\documentclass[aps,prb,preprint,groupedaddress]{revtex4-1}

\usepackage{graphicx,amsmath,amsfonts,mathrsfs}

\begin{document}


\title{Self-consistent linear response for the spin-orbit interaction related properties}


\author{I. V. Solovyev}
\email{SOLOVYEV.Igor@nims.go.jp}
\affiliation{Computational Materials Science Unit,
National Institute for Materials Science, 1-2-1 Sengen, Tsukuba,
Ibaraki 305-0047, Japan}


\date{\today}

\begin{abstract}
In many cases, the relativistic spin-orbit (SO) interaction can be regarded as a small perturbation
to the electronic structure of solids and treated using regular perturbation theory.
The major obstacle on this route comes from the fact that the SO interaction can also polarize the electron system
and produce some additional contributions to the perturbation theory expansion, which arise from the electron-electron
interactions in the same order of the SO coupling. In electronic structure calculations, it may even lead to necessity
to abandon the perturbation theory and return to the original self-consistently solution of Kohn-Sham-like equations
with the effective potential $\hat{v}$, incorporating simultaneously the effects of the electron-electron interactions and the
SO coupling, even though the latter is small.
In this work, we present the theory of self-consistent linear response (SCLR),
which allows us to get rid of numerical self-consistency and
formulate the last step fully analytically
in the first order of the SO coupling. This strategy is applied to the solution of
effective Hubbard-type model in unrestricted Hartree-Fock approximation.
The model itself is derived from the first-principles electronic structure calculations
in the basis of Wannier states close to the Fermi level and
is considered to be a good starting point for the analysis of magnetic properties of realistic transition-metal oxides.
We show that, by using $\hat{v}$, obtained in the SCLR theory,
one can successfully reproduce results of ordinary self-consistent calculations for
the orbital magnetization and other properties, which emerge in the first order of the SO coupling.
Particularly, SCLR appears to be extremely useful approach for
calculations of
antisymmetric Dzyaloshinskii-Moriya (DM) interactions based on the magnetic force theorem.
We argue that
only by using \textit{total} perturbation, one can make a
reliable estimate for the DM parameters.
Furthermore, due to the powerful $2n$$+$$1$ theorem,
the SCLR theory allows us to obtain the total energy change
up to the third order of the SO coupling, which can be used in calculations of
magnetic anisotropy of compounds with low crystal symmetry. The fruitfulness of this approach for the analysis of
complex magnetic structure
is illustrated on a number of example, including the quantitative description of
the spin canting in YTiO$_3$ and LaMnO$_3$, formation of the spin-spiral order in BiFeO$_3$, and the
magnetic inversion symmetry breaking in BiMnO$_3$, which gives rise to both
ferroelectric activity and DM interactions, responsible for the ferromagnetism.
\end{abstract}


\maketitle

\section{\label{sec:Intro} Introduction}

  The relativistic spin-orbit (SO) interaction is responsible for many interesting and important phenomena,
especially when
it comes into play with the magnetism. The typical examples in solids include the
orbital magnetization, magnetocrystalline anisotropy, magneto-optical activity, etc.\cite{LL,White,Yosida,Skomski}
The SO interaction can also contribute to the
magnetically induced
inversion symmetry breaking in multiferroics, which has attracted an enormous amount of attention
over the last decade.\cite{MF_review}

  In many applications, the SO interaction can be regarded to be small
and treated as a perturbation. For instance, the SO interaction
is usually much smaller than the
bandwidth in the vast majority of magnetic $3d$ compounds. The $3d$-level splitting,
which is often
caused by crystal distortions of a nonrelativistic origin, can be also substantially larger than the SO interaction.
The typical example of phenomena, which are treated perturbatively is
the orbital magnetization, appearing in the first order
of the SO coupling, and the uniaxial anisotropy, appearing in the second order.
In many cases, the perturbative treatment can be sufficient
from the viewpoint of numerical accuracy. It also provides a clear microscopic picture underlying
the considered phenomena.

  Nevertheless, the situation is severely complicated in the many-electron case: the SO coupling
will polarize the electron system, giving rise to the new terms, which arise from the electron-electron interactions and
additionally contribute to the perturbation theory expansion, formally in the same order of the SO coupling.
It can be well seen in the one-electron theories,
where the search of the magnetic ground state is reduced to the self-consistent
solution of ene-electron equations with the effective Kohn-Sham-like potential $\hat{v}$:\cite{WKohn}
the external perturbation $\delta \hat{v}^{ext}$ (in our case -- the SO interaction) produces some change of $\hat{v}$,
which can contribute to the magnetic properties in the same order of $\delta \hat{v}^{ext}$ and which should be found
self-consistently. Such a situation occurs, for instance, in the one-electron Hartree-Fock (HF) theory
or in the commonly used Kohn-Sham density-functional theory (DFT) and its various modifications.\cite{WKohn}

  The perturbative treatment of the SO interaction is widely used in electronic structure calculations based on DFT. For instance,
it was employed for the analysis of orbital magnetization and magnetocrystalline anisotropy energy,\cite{MAEexamples,PRB95}
antisymmetric Dzyaloshinskii-Moriya (DM) interactions,\cite{PRL96,DMexamples} etc. Most of such calculations
are supplemented with the frozen potential approximation (FPA), where the effect of the SO coupling is
evaluated for some fixed potential $\hat{v}$ (typically, obtained without SO coupling).
However, the validity of this strategy
crucially depends on the form of exchange-correlation potential. If it is smooth, as in
the local-spin-density approximation (LSDA), which is based on the theory of homogeneous electron gas,
the FPA works reasonably well: the LSDA potential is nearly spherical near atomic sites
and only weakly depends on the orbital variables. Another factor, which
justifies the use of the FPA is the additional spherical average of the potential inside
atomic spheres (the so-called atomic sphere approximation or ASA) in the linear muffin-tin orbitals method.\cite{LMTO}

  Nevertheless, it is commonly accepted today that the LSDA itself is inadequate for treating the
orbital magnetization and other SO interaction related properties, where one essential points is
to consider the explicit orbital dependence of the exchange-correlation energy, which is missing in LSDA.\cite{OPB,LDAU}
This is rather old problem,
which is also known as the treatment of the `orbital polarization'
or `population imbalance' in the electronic structure calculations.\cite{Terakura1984}
However, FPA is incompatible with the orbital polarization, which assumes an explicit and sometimes rather strong orbital
dependence of the exchange-correlation potential.
Thus, even though the SO interaction is small, in the latter case we are typically forced to run full-scale self-consistent
calculations with some orbital-dependent potential.

  Although we do not consider it in the present study, another important direction is the
refinement of the theory of orbital magnetization for extended periodic systems, which brings a number of new
and interesting ideas to this field.\cite{moderno,Nikolaev}

  In this article, we present the theory of self-consistent linear response (SCLR) for the SO interaction related properties.
This is also a perturbation theory with respect to the SO interaction, which allows us to get rid of numerical self-consistency
in the process of solution of one-electron Kohn-Sham-like equations, and to formulate this step analytically,
similar to the random-phase approximation (RPA) for the screened Coulomb interaction.\cite{scrRPA}
We will introduce our method for the unrestricted HF
solution of the low-energy model, derived from the
first-principles electronic structure calculations in the Wannier basis.\cite{review2008}
However, it can be generalized and applied for other types of
electronic structure calculations, such as LDA$+$$U$,\cite{AZA} hybrid functionals,\cite{Becke} etc.,
where the search of the magnetic ground state is
reduced to the solution of an auxiliary one-electron problem with the orbital-dependent
exchange-correlation potential. It should be noted that the idea itself is not new and was discussed in the
context of magnetic anisotropy calculations in the Anderson model already in 1960s.\cite{YosidaPTP}
However, it it practically forgotten today, despite its fruitfulness and importance
in the field of electronic structure calculations.
We will generalize this approach and derive an analytical expression for the self-consistent density matrix and
the effective potential $\hat{v}$ in the first order of the SO coupling, being affected by the
on-site Coulomb interactions and the
crystal field of an arbitrary form.
With this approach,
we will be able to access the behavior of the orbital magnetization
or any other property, which appears in the first
order of the SO coupling.
Then, due to variational character of the total energy and the powerful $2n$$+$$1$ theorem,\cite{2n+1}
the SCLR theory will enable us to obtain the energy change up to the third order of the SO coupling -- the property, which is
extremely useful for the analysis of magnetocrystalline anisotropy (MA) energy.
Moreover, by knowing self-consistent potential $\hat{v}$, one can derive all kind of properties
in the first order of the SO coupling from the single particle energies, by employing the
magnetic force theorem.\cite{MFT,JHeisenberg} We will use this approach in order to calculate parameters of
antisymmetric DM interactions (Ref.~\onlinecite{DM}) and, on the basis of these parameters,
to discuss the spin canting in
distorted transition-metal perovskite oxides. We will show that parameters of interatomic magnetic interactions,
obtained in such a way, reproduce nearly perfectly results of self-consistent electronic structure calculations
with the SO coupling. The ability of the proposed method will be illustrated on a number of example, including
the spin canting in YTiO$_3$ and LaMnO$_3$, formation of the spin-spiral order in BiFeO$_3$, and
magnetic control of ferroelectric (FE) polarization in multiferroic BiMnO$_3$, where the magnetic inversion symmetry breaking
gives rise not only to the FE activity, but also produces finite
DM interactions across the
inversion centers, which are responsible for the ferromagnetism.

  The rest of the article is organized as follows. In Sec.~\ref{sec:Method} we present our method:
a brief summary of the construction of effective low-energy model on the basis of first-principles electronic
structure calculations is given Sec.~\ref{sec:LEmodel},
the solution of this model using unrestricted HF approach is discussed in Sec.~\ref{sec:HF},
and the theory of self-consistent linear response is presented in Sec.~\ref{sec:LR}.
Sec.~\ref{sec:Appl} deals with applications of SCLR for calculation and analysis of local
spin and orbital magnetic moments (Sec.~\ref{sec:SL}), total energy (Sec.~\ref{sec:TE}), and
interatomic magnetic interactions (Sec.~\ref{sec:DM}).
In Sec.~\ref{sec:Results} we will employ SCLR for the analysis of spin caning in YTiO$_3$ (Sec.~\ref{sec:YTiO3})
and LaMnO$_3$ (Sec.~\ref{sec:LaMnO3}), spiral magnetic ordering in BiFeO$_3$ (Sec.~\ref{sec:BiMnO3}), and
the magnetic inversion symmetry breaking in BiMnO$_3$ (Sec.~\ref{sec:BiMnO3}).
Finally, in Sec. \ref{sec:Summary}, we will summarize our work. A brief summary of spin model
for the spiral magnetic phase, which can be realized in the rhombohedral $R3c$ structure of BiFeO$_3$, is presented in the Appendix.

\section{\label{sec:Method} Method}

\subsection{\label{sec:LEmodel} Effective low-energy model}

  The effective low-energy model is regarded as a bridge between first-principles electronic structure calculations
and the model Hamiltonian approach. With the proper construction, the model reproduces results of first-principles
calculations, at least on a semi-quantitative level. Moreover, the model allows us to treat the problem of electron correlations
beyond conventional approximations employed in the first-principles calculations. In this section, we briefly remind the reader
the main ideas of the construction of the model Hamiltonian. The details can be found in the review article (Ref.~\onlinecite{review2008})
and in previous publications.\cite{t2g,BiMnO3,JPSJ,NJP09}

  The model Hamiltonian,
\begin{equation}
\hat{\cal{H}}  =  \sum_{ij} \sum_{\alpha \beta}
t_{ij}^{\alpha \beta}\hat{c}^\dagger_{i\alpha}
\hat{c}^{\phantom{\dagger}}_{j\beta} +
  \frac{1}{2}
\sum_{i}  \sum_{\alpha \beta \gamma \delta} U^i_{\alpha \beta
\gamma \delta} \hat{c}^\dagger_{i\alpha} \hat{c}^\dagger_{i\gamma}
\hat{c}^{\phantom{\dagger}}_{i\beta}
\hat{c}^{\phantom{\dagger}}_{i\delta},
\label{eqn.ManyBodyH}
\end{equation}
is formulated
in the basis of Wannier orbitals $\{ \phi_{i \alpha} \}$,
which are constructed for the magnetically active bands near the Fermi level.
Here, each Greek symbol ($\alpha$, $\beta$, $\gamma$, or $\delta$) stands for the combination of
spin ($\sigma$$=$ $+$ or $-$) and orbital ($a$, $b$, $c$, or $d$) indices:
for instance, $\alpha \equiv (\sigma_\alpha, a)$, etc. Each lattice point $i$ ($j$) is specified by the position
$\boldsymbol{\tau}$ ($\boldsymbol{\tau}'$)
of the atomic site in the primitive cell and the lattice translation ${\bf R}$.

  The model is constructed
starting from the electronic band structure in the local-density approximation (LDA). The first step is the
construction of localized Wannier basis for the magnetically active bands.\cite{WannierRevModPhys}
Each basis orbital
$\phi_{\boldsymbol{\tau} \alpha} ({\bf r}-{\bf R})$ is labeled by the combined index $\alpha$ and centered around
some lattice point $(\boldsymbol{\tau}$$+$${\bf R})$.
In our case, the Wannier function were generated using the
projector-operator technique (Ref.~\onlinecite{review2008})
and orthonormal basis orbitals of the LMTO method (Ref.~\onlinecite{LMTO}) as the trial wave functions.
As the LMTO basis functions are already well localized, typically such procedure allows us to generate
well localized Wannier functions.
Then, the one-electron part of the model
is identified with the matrix elements of LDA Hamiltonian ($\cal{H}_{\rm LDA}$) in the Wannier basis:
$t^{\alpha \beta}_{\boldsymbol{\tau}, \boldsymbol{\tau}'+{\bf R}} =
\langle \phi_{\boldsymbol{\tau} \alpha} ({\bf r})| \cal{H}_{\rm LDA} | \phi_{\boldsymbol{\tau}' \beta} ({\bf r}-{\bf R}) \rangle$.
Since the Wannier basis is complete in the low-energy part of the spectrum, the construction is exact in the sense that
the band structure, obtained from $t^{\alpha \beta}_{\boldsymbol{\tau}, \boldsymbol{\tau}'+\bf R}$,
exactly
coincides with the one of LDA.

  The Wannier basis and the one-electron parameters $t^{\alpha \beta}_{\boldsymbol{\tau}, \boldsymbol{\tau}'+{\bf R}}$ were
first computed without SO interaction. In this case, the matrix elements do not depend on the spin indices:
$t^{\alpha \beta}_{\boldsymbol{\tau}, \boldsymbol{\tau}'+{\bf R}} =
t^{ab}_{\boldsymbol{\tau}, \boldsymbol{\tau}'+{\bf R}} \delta_{\sigma_\alpha \sigma_\beta}$.
After that, matrix elements of the SO interaction were calculated at each atomic site in the same
`nonrelativistic' basis of Wannier orbitals:
$\langle \phi_{\boldsymbol{\tau} \alpha} ({\bf r})| \Delta \cal{H}_{\rm SO} | \phi_{\boldsymbol{\tau} \beta} \rangle$,
as explained in Ref.~\onlinecite{review2008}. In the regular HF calculations, these matrix elements are combined with the previously computed
`nonrelativistic' $t^{\alpha \beta}_{\boldsymbol{\tau}, \boldsymbol{\tau}'+{\bf R}}$, while in SCLR this part is treated
as the external perturbation $\delta \hat{v}^{ext}$.

  Matrix elements of screened Coulomb interactions at some atomic site $\boldsymbol{\tau}$ can be also calculated in the Wannier basis.
For the purposes of our work, it is convenient to adopt the following notations:
\begin{equation}
U_{\alpha \beta \gamma \delta}^{\boldsymbol{\tau}} = \int d {\bf r} \int d {\bf r}'
\phi_{\boldsymbol{\tau} \alpha}^* ({\bf r})
\phi_{\boldsymbol{\tau} \beta} ({\bf r}) v_{\rm scr}({\bf r},{\bf r}')
\phi_{\boldsymbol{\tau} \gamma}^* ({\bf r}') \phi_{\boldsymbol{\tau} \delta} ({\bf r}').
\label{eqn:scrU}
\end{equation}
The screened Coulomb interaction
$v_{\rm scr}({\bf r},{\bf r}')$ can be computed by employing the constrained RPA technique.\cite{Ferdi04}
In this case, $v_{\rm scr}({\bf r},{\bf r}')$ does not depend on spin variables and
$U_{\alpha \beta \gamma \delta}^{\boldsymbol{\tau}} = U_{abcd}^{\boldsymbol{\tau}} \,
\delta_{\sigma_\alpha \sigma_\beta} \delta_{\sigma_\gamma \sigma_\delta}$.
Since the constrained RPA technique is very time consuming, we apply additional approximations, which were discussed
in Ref.~\onlinecite{review2008}. Namely, first we evaluate the screened Coulomb and exchange interactions between
atomic $3d$ orbitals, using fast and more suitable for these purposes constrained LDA technique. After that,
we consider additional channel of screening caused by the $3d \rightarrow 3d$ transitions
in the framework of constrained RPA technique and projecting corresponding
polarization function onto the $3d$ orbitals. The so obtained parameters of Coulomb interactions
are well consistent with results of full-scale constrained RPA calculations.\cite{cRPA}

  Parameters of model Hamiltonian for YTiO$_3$, LaMnO$_3$, and BiMnO$_3$ were already discussed in
the previous publications
(Refs.~\onlinecite{review2008,t2g,NJP09} for YTiO$_3$, Ref.~\onlinecite{JPSJ} for LaMnO$_3$, and
Ref.~\onlinecite{BiMnO3} for BiMnO$_3$).
Therefore, we will not consider them here again. The parameters for BiFeO$_3$ are summarized in supplemental materials.\cite{SM}
The model was constructed for the magnetically active $t_{2g}$ bands in the case of YTiO$_3$, and all $3d$ bands
in the case of LaMnO$_3$, BiMnO$_3$, and BiFeO$_3$.

\subsection{\label{sec:HF} Unrestricted Hartree-Fock approach}

  One-electron HF equations for periodic Hubbard model
can be conveniently formulated in the matrix form, in each point of the Brillouin zone (BZ):
\begin{equation}
\hat{{\cal H}}_{\rm HF}({\bf k}) |C_{\nu {\bf k}} \rangle =
\varepsilon_{\nu {\bf k}} |C_{\nu {\bf k}} \rangle,
\label{eqn:HF}
\end{equation}
where the matrix $\hat{{\cal H}}_{\rm HF}({\bf k}) = [\hat{t}({\bf k}) + \hat{v}]$ and the column vector $|C_{\nu {\bf k}} \rangle$
are specified by three types of indices: spin,
orbital, and the position of the
atomic site in the primitive cell. The band index $\nu$ also includes the
information about the spin of electron.

  $\hat{t}({\bf k})$ is the supermatrix, which is composed from the matrices
$$
\hat{t}_{\boldsymbol{\tau} \boldsymbol{\tau}'}({\bf k}) = \sum_{\bf R}
\hat{t}_{\boldsymbol{\tau}, \boldsymbol{\tau}'+\bf R}
e^{i{\bf k}\cdot(\boldsymbol{\tau}' - \boldsymbol{\tau} + {\bf R})},
$$
for each pair of atomic sites $\boldsymbol{\tau}$ and $\boldsymbol{\tau}'$,
where $\hat{t}_{\boldsymbol{\tau}, \boldsymbol{\tau}'+\bf R} \equiv [ t^{\alpha \beta}_{\boldsymbol{\tau}, \boldsymbol{\tau}'+\bf R}  ]$.

  $\hat{v}$ is the self-consistent HF potential, which is diagonal with respect to the site indices. For each $\boldsymbol{\tau}$,
it can be found using site-diagonal elements
of screened Coulomb ($U_{abcd}$) and
the density matrix $\hat{n} = [ n_{ab}^{\sigma \sigma'} ]$:
\begin{equation}
n_{ab}^{\sigma \sigma'} = \sum_{\nu}^{\rm occ} \sum_{\bf k}^{\rm BZ}
\left( C_{\nu {\bf k}}^{a \sigma} \right)^* C_{\nu {\bf k}}^{b \sigma'},
\label{eqn:DM}
\end{equation}
where $C_{\nu {\bf k}}^{a \sigma}$ and $C_{\nu {\bf k}}^{b \sigma'}$ are the elements of the vector $|C_{\nu {\bf k}} \rangle$,
the first summation runs over occupied (${\rm occ}$)
states and the second summation -- over first BZ.
Using the notations adopted in Eq.~(\ref{eqn:scrU}),
diagonal and non-diagonal matrix elements of $\hat{v}$ with respect to the spin indices are given by
\begin{equation}
v_{ab}^{\sigma \sigma} = \sum_{cd} \left\{
(U_{abcd}-J_{abcd}) n_{cd}^{\sigma \sigma} + U_{abcd} n_{cd}^{\bar{\sigma} \bar{\sigma}} \right\}
\label{eqn:HFpot1}
\end{equation}
and
\begin{equation}
v_{ab}^{\sigma \bar{\sigma}} = - \sum_{cd}
J_{abcd} n_{cd}^{\bar{\sigma} \sigma},
\label{eqn:HFpot2}
\end{equation}
respectively,
where $\bar{\sigma} = -$$\sigma$ and $J_{abcd} = U_{adcb}$.
The non-diagonal elements $v_{ab}^{\sigma \bar{\sigma}}$ can arise from the SO interaction and/or noncollinear
magnetic alignment.
Eqs.~(\ref{eqn:HF})-(\ref{eqn:HFpot2})
are known as self-consistent equations of unrestricted HF method. All quantities
in these equations can depend on the site-index $\boldsymbol{\tau}$, which we drop for simplicity, unless it
is specified otherwise.

\subsection{\label{sec:LR} Self-consistent linear response}

  Let us start with a collinear spin structure without SO coupling. Then, $\hat{t}({\bf k})$ does not depend on
spin indices, $\hat{v}$ can be chosen to be diagonal with respect to the spin indices, so as the full Hamiltonian of the
HF method:
\begin{equation}
\hat{{\cal H}}_{\rm HF} = \left(
\begin{array}{cc}
\hat{{\cal H}}_{\rm HF}^+ & 0 \\
0 & \hat{{\cal H}}_{\rm HF}^- \\
\end{array}
\right).
\label{eqn:HHF}
\end{equation}
In this case, each state $\nu$ can be characterized by
its spin $\sigma$ and the remaining band index $m$, numbering the bands for each spin: $\nu \equiv (\sigma m)$.

  Next, let us consider the external perturbation $\delta \hat{v}^{ext}$, which we assume to be periodic and diagonal with
respect to the site indices $\boldsymbol{\tau}$. It could be the SO interaction or any other interaction, obeying this property.
First, we are interested in the change of the density matrix, which is cause by $\delta \hat{v}^{ext}$.
In the framework of the linear response theory, this change can be identically written as
$$
\delta \hat{n} = \boldsymbol{\cal R} \delta \hat{v}^{ext},
$$
where elements of the tensor $\boldsymbol{\cal R} \equiv [ {\cal R}^{\sigma \sigma'}_{abcd} ]$
can be obtained by applying the regular perturbation theory to the HF eigenvectors.
Then, Eq.~(\ref{eqn:DM}) yields in the first
order of $\delta \hat{v}^{ext}$:
$$
{\cal R}^{\sigma \sigma'}_{abcd} = \sum_m^{\rm occ} \sum_l^{\rm unocc} \sum_{\bf k}^{\rm BZ}
\left\{
\frac{(C_{m {\bf k}}^{a \sigma})^* C_{l {\bf k}}^{b \sigma'} (C_{l {\bf k}}^{c \sigma'})^* C_{m {\bf k}}^{d \sigma}}
{\varepsilon_{m \sigma {\bf k}} - \varepsilon_{l \sigma' {\bf k}}}
+
\frac{(C_{l {\bf k}}^{a \sigma})^* C_{m {\bf k}}^{b \sigma'} (C_{m {\bf k}}^{c \sigma'})^* C_{l {\bf k}}^{d \sigma} }
{\varepsilon_{m \sigma' {\bf k}} - \varepsilon_{l \sigma {\bf k}}}
\right\},
$$
where the first and second summation runs over the occupied (${\rm occ}$) and unoccupied (${\rm unocc}$) states, respectively,
and the ${\bf k}$-summation -- over the first BZ. As was explained above, we drop for simplicity the site-indices
$\boldsymbol{\tau}$ and $\boldsymbol{\tau}'$.
However, it should be understood
that the orbitals $a$ and $b$ belong to one site (say, $\boldsymbol{\tau}$), while the orbitals $c$ and $d$ can
belong to another site (say, $\boldsymbol{\tau}'$), which is generally different from $\boldsymbol{\tau}$.
Then, the tensor multiplication $\boldsymbol{\cal R} \delta \hat{v}^{ext}$
implies the summation over the indices $c$, $d$ (and $\boldsymbol{\tau}'$), while the indices
$a$, $b$ (and $\boldsymbol{\tau}$) specify the matrix element $\delta n_{ab}^{\sigma \sigma'}$ of the density matrix.
The relationship between spin indices of $\delta \hat{n}$
and $\delta \hat{v}^{ext}$ can be best understood by introducing the vector
$$
\vec{\delta n} = \left(
\begin{array}{c}
\delta \hat{n}^{++} \\
\delta \hat{n}^{-+} \\
\delta \hat{n}^{+-} \\
\delta \hat{n}^{--} \\
\end{array}
\right)
$$
(and a similar vector for other matrices such as $\vec{\delta v}^{ext}$, etc.),
where each $\delta \hat{n}^{\sigma \sigma'}$ is a matrix in the subspace of orbital indices.
Then, $\vec{\delta n}$ and $\vec{\delta v}^{ext}$ will be related by the matrix equation
$$
\vec{\delta n} = {\mathbb R} \vec{\delta v}^{ext},
$$
where
$$
{\mathbb R} =
\left(
\begin{array}{cccc}
\boldsymbol{\cal R}^{++} &             0            &          0               &             0            \\
            0            &             0            & \boldsymbol{\cal R}^{-+} &             0            \\
            0            & \boldsymbol{\cal R}^{+-} &          0               &             0            \\
            0            &            0             &          0               & \boldsymbol{\cal R}^{--} \\
\end{array}
\right)
$$
and each sub-block $\boldsymbol{\cal R}^{\sigma \sigma'}$ is composed from the elements ${\cal R}^{\sigma \sigma'}_{abcd}$
with the given spin indices $\sigma$ and $\sigma'$. Then, knowing $\vec{\delta n}$ and using Eqs.~(\ref{eqn:HFpot1})-(\ref{eqn:HFpot2}),
one can find the change of the HF potential. For these purposes, it is
convenient to introduce the matrix
$$
{\mathbb U} =
\left(
\begin{array}{cccc}
\boldsymbol{\cal U}-\boldsymbol{\cal J} &             0            &          0               &                     \boldsymbol{\cal U} \\
            0                           &             0            & -\boldsymbol{\cal J}     &                            0            \\
            0                           & -\boldsymbol{\cal J}     &          0               &             0                           \\
 \boldsymbol{\cal U}                    &            0             &          0               & \boldsymbol{\cal U}-\boldsymbol{\cal J} \\
\end{array}
\right),
$$
where again each sub-block is composed from $U_{abcd}$ and (or) $J_{abcd}$. Then, Eqs.~(\ref{eqn:HFpot1})-(\ref{eqn:HFpot2}) can be written
in the compact form:
$$
\vec{v} = {\mathbb U} \vec{n}.
$$
Therefore, the change of the HF potential will be given by
$$
\vec{\delta v} = {\mathbb U} {\mathbb R} \vec{\delta v}^{ext},
$$
where the matrix multiplication ${\mathbb U} {\mathbb R}$ also implies the summation over two orbital indices. After that
$\vec{\delta v}$ can be combined with $\vec{\delta v}^{ext}$ and the problem can be solved self-consistently,
similar to calculations of screened Coulomb interaction in RPA,\cite{scrRPA}
where the total perturbation
($\vec{\delta v}^p = \vec{\delta v}^{ext} + \vec{\delta v}$) on the input of \textit{n}-\textit{th} iteration is related to the previous one by the condition
$$
(\vec{\delta v}^p)^{(\textit{n})} = {\mathbb U} {\mathbb R} (\vec{\delta v}^p)^{(\textit{n}-1)} + \vec{\delta v}^{ext}.
$$
This yields self-consistent solution for $\vec{\delta v}^p$, which is valid in the first order of $\vec{\delta v}^{ext}$:
\begin{equation}
\vec{\delta v}^p = \left[ 1-{\mathbb U} {\mathbb R} \right]^{-1} \vec{\delta v}^{ext}.
\label{eqn:SCpert}
\end{equation}
Then, the change of the density matrix and the HF potential $\vec{\delta v} = \vec{\delta v}^p - \vec{\delta v}^{ext}$
in the first order of $\delta \hat{v}^{ext}$ will be given by
\begin{equation}
\vec{\delta n} = {\mathbb R} \vec{\delta v}^p
\label{eqn:SCden}
\end{equation}
and
\begin{equation}
\vec{\delta v} = {\mathbb U} {\mathbb R} \vec{\delta v}^p,
\label{eqn:SCpot}
\end{equation}
respectively.
By knowing $\vec{\delta v}$ and $\vec{\delta n}$,
one should be able to calculate all possible properties in the first order of $\delta \hat{v}^{ext}$, and the
total energy -- up to the third order in $\delta \hat{v}^{ext}$.\cite{2n+1} Again, tracing an analogy with RPA,
$\vec{\delta v}^p$, given by Eq.~(\ref{eqn:SCpert}), can be regarded as the `screened' SO interaction,
corresponding to the `bare' interaction $\vec{\delta v}^{ext}$.

  Eqs.~(\ref{eqn:SCpert})-(\ref{eqn:SCpot}) are subjected to some instabilities, which are
signalled by the poles of $\left[ 1-{\mathbb U} {\mathbb R} \right]^{-1}$. Among them, there is a
trivial instability towards uniform rotation of the spin system as the whole, which we have to remove.
For these purposes we constrain the matrix elements of $\boldsymbol{\cal R}$ so that, at each iteration,
the `corrected' tensor $\tilde{\boldsymbol{\cal R}}$ would generate the density matrix
$\delta \hat{\tilde{n}}$, satisfying the following condition. Let
$\boldsymbol{\mu}^0_{\boldsymbol{\tau}} = {\rm Tr} [ \hat{\boldsymbol{\sigma}} \hat{n}_{\boldsymbol{\tau}} ]$
be the spin magnetic moment at the sites ${\boldsymbol{\tau}}$
without $\delta \hat{v}^{ext}$ (in the considered geometry, $\boldsymbol{\mu}^0_{\boldsymbol{\tau}} || z$), and
$\delta \boldsymbol{\mu}^S_{\boldsymbol{\tau}} = {\rm Tr} [ \hat{\boldsymbol{\sigma}} \delta \hat{n}_{\boldsymbol{\tau}} ]$
is the perturbation caused by $\delta \hat{v}^{ext}$, where ${\rm Tr}$ is the trace running over spin and orbital indices,
and $\hat{\boldsymbol{\sigma}}$ is the vector of Pauli matrices.
The magnetic moments throughout this paper are quoted in units of Bohr magneton, $\mu_B$.
Moreover, here we recall the explicit dependence of $\hat{n}$ and $\delta \hat{n}$ on the site-indices $\boldsymbol{\tau}$.
Then, the spin system will experience the uniform rotation if
$\sum_{\boldsymbol{\tau}} [ \boldsymbol{\mu}^0_{\boldsymbol{\tau}} \times \delta \boldsymbol{\mu}^S_{\boldsymbol{\tau}} ] \ne 0$.
Therefore, we define the constrained density matrix as
$$
\delta \hat{\tilde{n}}_{\boldsymbol{\tau}} = \delta \hat{n}_{\boldsymbol{\tau}} - \boldsymbol{\lambda} \cdot
[\boldsymbol{\mu}^0_{\boldsymbol{\tau}} \times \hat{\boldsymbol{\sigma}}],
$$
and find $\boldsymbol{\lambda}$ from the condition
$$
\sum_{\boldsymbol{\tau}} [ \boldsymbol{\mu}^0_{\boldsymbol{\tau}} \times \delta \tilde{\boldsymbol{\mu}}^S_{\boldsymbol{\tau}} ] = 0,
$$
where $\delta \tilde{\boldsymbol{\mu}}^S_{\boldsymbol{\tau}} =
{\rm Tr} [ \hat{\boldsymbol{\sigma}} \delta \hat{\tilde{n}}_{\boldsymbol{\tau}} ]$.

  Finally, the above strategy was considered for the case where the spin quantization axis without $\delta \hat{v}^{ext}$
is parallel to $z$. However, it can be easily generalized for an arbitrary direction
$\boldsymbol{e} = (\cos \varphi \sin \vartheta, \sin \varphi \sin \vartheta, \cos \vartheta)$
of the spin quantization axis,
by applying the matrices of spin rotations to $\delta \hat{v}^{ext}$:
$$
\delta \hat{v}^{ext} \rightarrow \hat{\mathscr{U}}(\vartheta,\varphi) \delta \hat{v}^{ext} \hat{\mathscr{U}}^\dagger (\vartheta,\varphi),
$$
and, then, to the obtained matrices $\delta \hat{n}$ and $\delta \hat{v}$.

\section{\label{sec:Appl} Applications}

  In this section, we consider several useful applications of SCLR.

\subsection{\label{sec:SL} Spin and Orbital Magnetic Moments}

  In most of the cases, the orbital magnetic moment, $\boldsymbol{\mu}^L_{\boldsymbol{\tau}}$,
is induced by the SO coupling ($\delta \hat{v}^{ext}$) and additionally
enhanced by electron-electron interactions in the system.\cite{OPB,LDAU}
$\boldsymbol{\mu}^L_{\boldsymbol{\tau}}$ can be often well described in the
first order perturbation theory with respect to the SO coupling.
The main obstacle, however, was how to incorporate the effect of electron-electron interactions
into the perturbation theory. This problem is perfectly solved by SCLR, which provides
$\delta \hat{n}$ in the first order of the SO coupling and this $\delta \hat{n}$
already includes the effects of the electron-electron interactions
in the same first order of the SO coupling. Then,
the quantity
$$
\boldsymbol{\mu}^L_{\boldsymbol{\tau}} = {\rm Tr} [ \hat{\boldsymbol{L}} \delta \hat{n}_{\boldsymbol{\tau}} ]
$$
(where $\hat{\boldsymbol{L}}$ is the angular momentum operator in the Wannier basis)
should provide a good estimate for the orbital magnetization, at least for its local part.\cite{Nikolaev}

  The spin magnetization can be also estimated in the first order of the SO coupling as
$$
\boldsymbol{\mu}^S_{\boldsymbol{\tau}} = \boldsymbol{\mu}^0_{\boldsymbol{\tau}} + \delta \boldsymbol{\mu}^S_{\boldsymbol{\tau}},
$$
where $\boldsymbol{\mu}^0_{\boldsymbol{\tau}} = {\rm Tr} [ \hat{\boldsymbol{\sigma}} \hat{n}_{\boldsymbol{\tau}} ]$
and $\delta \boldsymbol{\mu}^0_{\boldsymbol{\tau}} = {\rm Tr} [ \hat{\boldsymbol{\sigma}} \delta \hat{n}_{\boldsymbol{\tau}} ]$.
However, $\boldsymbol{\mu}^S_{\boldsymbol{\tau}}$ can be also sensitive to some
higher-order effects of the SO coupling. For example, the directions of $\boldsymbol{\mu}^S_{\boldsymbol{\tau}}$ are known
to be affected by the single-ion (SI) anisotropy, which emerges in the second order of the SO coupling and
is formally beyond the accessibility of SCLR.
This is the main reason why SCLR provides much better estimate for the
orbital magnetization than for the spin one.
As we will see below, SCLR does allow us to consider some higher-order effects for the total energy.
However, the corresponding change of the density matrix, $\delta \hat{n}$, is essentially limited by the
first order of the SO coupling. This constitutes the main limitation of SCLR for treating the spin magnetization.

\subsection{\label{sec:TE} Total energy}

  The total energy in the unrestricted HF approach can be written in the compact form as
\begin{equation}
E = \sum_{\nu}^{\rm occ} \sum_{\bf k}^{\rm BZ} \varepsilon_{\nu {\bf k}} +
{\rm Tr} \left\{ \frac{1}{2} \vec{n}^{\,T}{\mathbb U}\vec{n} - \vec{n}^{\,T} \vec{v} \right\},
\label{eqn:totalE}
\end{equation}
where the first term is the sum of single-particle energies ($E_{\rm sp}$) and the second one is the double-counting correction ($E_{\rm dc}$).
Here, we continue to use notations of Sec.~\ref{sec:LR} and, in the second term, drop for simplicity the summation
over site indices in the primitive cell. Moreover, $\vec{n}^{\,T}$ denotes the row vector of the form:
$$
\vec{n}^{\,T} = \left( \hat{n}^{++} \, \, \hat{n}^{-+} \, \, \hat{n}^{+-} \, \, \hat{n}^{--} \right).
$$

  Due to the time-reversal symmetry of unperturbed Hamiltonian $\hat{{\cal H}}_{\rm HF}^{\sigma}$ for each projection of spin,
there will be no
first-order contribution of the SO interaction to $E$.

  Then, by knowing $\vec{\delta v}$ and $\vec{\delta n}$ in the first order of the SO coupling, one can easily
find the correction to the total energy in the second order. The change of the single-particle energies
can be obtained in the second order perturbation theory, which yields
\begin{equation}
E_{\rm sp} = \frac{1}{2} {\rm Tr} \left\{ ( \vec{\delta v}^p )^T {\mathbb R} \vec{\delta v}^p \right\}.
\label{eqn:dEsp}
\end{equation}
Using Eq.~(\ref{eqn:SCden}),
it can be further transformed to $E_{\rm sp} = \frac{1}{2} {\rm Tr} \left\{ \vec{\delta n}^T \vec{\delta v}^p  \right\}$.
The change of the double-counting energy can be written as $E_{\rm dc} = -$$\frac{1}{2} {\rm Tr} \left\{ \vec{\delta n}^{\,T} \vec{\delta v} \right\}$.
By combining these two contributions and noting that $\vec{\delta v}^p = \vec{\delta v}^{exp} + \vec{\delta v}$, one
obtains the following expression for the change of the total energy in the second order of the SO interaction:
\begin{equation}
\delta E = \frac{1}{2} {\rm Tr} \left\{ \vec{\delta n}^T \vec{\delta v}^{ext} \right\}.
\label{eqn:dEt}
\end{equation}
Using Eq.~(\ref{eqn:SCden}), it can be also transformed to
\begin{equation}
\delta E = \frac{1}{2} {\rm Tr} \left\{ ( \vec{\delta v}^p )^T {\mathbb R} \vec{\delta v}^{ext} \right\}.
\label{eqn:dEtotal}
\end{equation}
It is also important that the second-order contribution to the potential itself is exactly canceled out between
the single-particle and double-counting terms. Thus, this contribution need not be considered.
It also justifies the use of magnetic force theorem, which, in the leading (first) order of the perturbation theory,
allows us to replace the total energy change
by the change of the single-particle energies.\cite{MFT,JHeisenberg}

  The second order perturbation theory for the total energy is already very useful for the analysis of MA energy and,
as will be discussed in Sec.~\ref{sec:Results},
typically reproduces results of fully self-consistent non-perturbative HF calculations within 10\% error. Nevertheless, the SCLR theory
allows us to make one step further. The reason for it is the variational character of the total energy, which
gives rise to the powerful $2n$$+$$1$ theorem.\cite{2n+1} It states that by knowing wavefunctions
(and, therefore, the density matrix) up to order $n$ with respect to some perturbation, one should be able to calculate
the change of the total energy up to order $2n$$+$$1$. Particularly, by knowing $\delta \hat{n}$ in the first order
of the SO interactions, one should be able to evaluate $\delta E$ in the third order. For these purposes, we used very
straightforward procedure: we took $\delta \hat{v}^p$, obtained in SCLR; calculated new sets of $\{ \varepsilon_{\nu {\bf k}} \}$
and $\{ |C_{\nu {\bf k}} \rangle \}$ for the
potential $\hat{v}$$+$$\delta \hat{v}^p$, using
Eq.~(\ref{eqn:HF}); found new $\hat{n}$, using Eq.~(\ref{eqn:DM}); and, then, evaluated the total energy,
using Eq.~(\ref{eqn:totalE}). This procedure gives us nearly perfect agreement with results of
fully self-consistent non-perturbative HF calculations
for the MA energy.

\subsection{\label{sec:DM} Spin model and Dzyaloshinskii-Moriya interactions}

  In this section, we investigate abilities of mapping of the total energies, obtained in the HF approximation
for the electronic model (\ref{eqn.ManyBodyH}),
onto the classical spin model
\begin{equation}
E = - \sum_{i > j} J_{ij} \boldsymbol{e}_i \cdot \boldsymbol{e}_j +
\sum_{i > j} \boldsymbol{d}_{ij} \cdot [\boldsymbol{e}_i \times \boldsymbol{e}_j] +
\sum_{i \ge j} \boldsymbol{e}_i \cdot \tensor{\tau}_{ij} \boldsymbol{e}_j,
\label{eqn:spinE}
\end{equation}
where $\boldsymbol{e}_i$ is the \textit{direction} of spin at the site $i$. Here, the first term stands
for isotropic Heisenberg interactions ($E_{\rm H}$), the second term -- for antisymmetric
Dzyaloshinskii-Moriya (DM) interactions ($E_{\rm DM}$), and the third one -- for symmetric anisotropic interaction,
where $i$$=$$j$ corresponds to the
SI anisotropy energy ($E_{\rm SI}$).
The leading contribution of the SO coupling to $J_{ij}$, $\boldsymbol{d}_{ij}$, and $\tensor{\tau}_{ij}$
is of the zeroth, first, and second order, respectively.

  Since SCLR allows us to evaluate the one-electron potential $\delta \hat{v}$ in the first order of the SO coupling,
we should be able to obtain all kind of the ground-state properties, in the same first order of the SO coupling, by
continuing to stay in the frameworks of the one-electron theory and applying the magnetic force theorem.\cite{MFT}
In the present section, we will focus on the force
$\boldsymbol{f}_i = -$$\partial E/\partial \boldsymbol{e}_i$, rotating the spin at the site $i$.
For the spin model (\ref{eqn:spinE}), it can be written in the form:
$$
\boldsymbol{f}_i = \sum_j \boldsymbol{f}_i^j,
$$
where
$$
\boldsymbol{f}_i^j = [ \boldsymbol{d}_{ij} \times \boldsymbol{e}_j] + J_{ij} \boldsymbol{e}_j
$$
is the force, acting on the spin $i$ from the spin $j$. Since we are interested only in the first-order effects with respect to the
SO coupling, we drop here the contribution of $\tensor{\tau}_{ij}$. Then, if $\boldsymbol{e}_j^0$ is the
direction of spin without SO coupling and $\delta \boldsymbol{e}_j = \boldsymbol{e}_j - \boldsymbol{e}_j^0$
is a correction, one can write, in the first order of the SO interaction:
$$
\boldsymbol{f}_i^j = [ \boldsymbol{d}_{ij} \times \boldsymbol{e}_j^0] + J_{ij} \delta \boldsymbol{e}_j,
$$
where the first term describes the rotation of spin at the site $i$ by the DM interaction $\boldsymbol{d}_{ij}$,
and the second term is due to rotation of spin ($\delta \boldsymbol{e}_j$) at the site $j$.
For the sake of clarity, let us consider the FM alignment where $\boldsymbol{e}_j^0 = (0,0,1) \equiv \boldsymbol{e}^0$ at all magnetic sites.
The generalization to the AFM case is straightforward, but a little bit more cumbersome.
For our purposes, it is convenient to consider the following antisymmetric construction:
\begin{equation}
\frac{1}{2} \left( \boldsymbol{f}_i^j - \boldsymbol{f}_j^i \right) =
[ \boldsymbol{d}_{ij} \times \boldsymbol{e}^0] + \frac{1}{2} J_{ij} \left( \delta \boldsymbol{e}_j - \delta \boldsymbol{e}_i \right).
\label{eqn:spinf}
\end{equation}

  Our next goal is to calculate a similar quantity for the electronic model (\ref{eqn:HF})
and to make a mapping on Eq.~(\ref{eqn:spinf}). For these purposes, it is convenient to use Lloyd's formula,
which was extensively used in the multiple scattering theory.\cite{Lloyd}
The formula states that any change of the single-particle energy in the second order of perturbation theory
can be expressed as a sum of pairwise interactions:
$$
\delta E_{\rm sp} = \sum_{i \ge j} \delta E_{ij},
$$
where
\begin{equation}
(1+\delta_{ij})
\delta E_{ij} = -\frac{1}{\pi} {\rm Im} \int_{- \infty}^{\varepsilon_F} d \varepsilon {\rm Tr}
\left\{ \hat{\cal G}_{ij}(\varepsilon) \delta \hat{v}_j \hat{\cal G}_{ji}(\varepsilon) \delta \hat{v}_i \right\},
\label{eqn:pairI}
\end{equation}
$\delta_{ij}$ is the Kronecker delta, and $\varepsilon_F$ is the Fermi level. Then, in our case,
$$
\hat{\cal G}_{ij}(\varepsilon) = \left(
\begin{array}{cc}
\hat{\cal G}_{ij}^+(\varepsilon) & 0 \\
0 & \hat{\cal G}_{ij}^-(\varepsilon) \\
\end{array}
\right)
$$
is Green's function for the Hamiltonian (\ref{eqn:HHF})
without
SO coupling, after the Fourier transformation to the real space:
$$
\hat{\cal G}_{ij}^{\sigma}(\varepsilon) = \left[ \varepsilon - \hat{{\cal H}}_{\rm HF}^{\sigma} + i0^+ \right]_{ij}^{-1},
$$
and the perturbation of the
potential at the site $i$ has two parts: $\delta \hat{v}_i \equiv \delta \hat{v}_i^{p} + \delta \hat{v}_i^{r}$.
The first one ($\delta \hat{v}_i^{p}$) is the total perturbation (external plus the change of the HF potential),
caused by the SO interaction, which is evaluated
using Eq.~(\ref{eqn:SCpert})
in the framework of the
SCLR theory. The SO interaction may be coupled to the rotation of the exchange spin field,
which is described by the
second term:
$$
\delta \hat{v}_i^{r} = \hat{b}_i \delta \boldsymbol{e}_i \cdot \hat{\boldsymbol{\sigma}},
$$
where $\hat{b}_i = \frac{1}{2}(\hat{v}_i^+$$-$$\hat{v}_i^-)$ is the exchange field at the siet $i$,
calculated without SO coupling. Moreover, it is understood that, if $\boldsymbol{e}_i^0$ is parallel
to the $z$ axis, $\delta \boldsymbol{e}_i$ lies in the $xy$ plane.
We would like to emphasize that Eq.~(\ref{eqn:pairI}) is nothing but the second order perturbation theory
for the single-particle energy, which is equivalent to formulation in terms of the response tensor,
Eq.~(\ref{eqn:dEsp}), except that Eq.~(\ref{eqn:pairI}) deals with a more general type of perturbation,
which is not necessary periodic.

  Then, by retaining only $\delta \hat{v}^{r}$ at the sites $i$ and $j$, Eq.~(\ref{eqn:pairI}) can be mapped onto
Heisenberg model:\cite{JHeisenberg}
$$
\delta E_{ij} = -J_{ij} \delta \boldsymbol{e}_i \cdot \delta \boldsymbol{e}_j,
$$
where
\begin{equation}
J_{ij} = \frac{2}{\pi} {\rm Im} \int_{- \infty}^{\varepsilon_F} d \varepsilon {\rm Tr}_L
\left\{ \hat{\cal G}_{ij}^+(\varepsilon) \hat{b}_j \hat{\cal G}_{ji}^-(\varepsilon) \hat{b}_i \right\}
\label{eqn:Jij}
\end{equation}
and ${\rm Tr}_L$ is the trace over orbital indices.

  In the same way, one can consider the mixed perturbation, where $\delta \hat{v}^{r}$ occurs at the site $i$,
and $\delta \hat{v}^{p}$ -- at the site $j$ (and vice versa). Noting that
$\frac{\partial}{\partial \boldsymbol{e}_i} \delta \hat{v}_i^{r} = \hat{b}_i \hat{\boldsymbol{\sigma}}$ and
$\frac{\partial}{\partial \boldsymbol{e}_i} \delta \hat{v}_i^{p}$ does not contribute to the forces in the first order
of the SO interaction, one obtains the following expression:
\begin{equation}
\frac{1}{2} \left( \boldsymbol{f}_i^j - \boldsymbol{f}_j^i \right) =
\frac{1}{\pi} {\rm Im} \int_{- \infty}^{\varepsilon_F} d \varepsilon {\rm Tr} \left\{
\hat{\boldsymbol{\sigma}} \left(
\hat{\cal G}_{ij}(\varepsilon) \delta \hat{v}_j^{p} \hat{\cal G}_{ji}(\varepsilon) \hat{b}_i -
\hat{\cal G}_{ij}(\varepsilon) \hat{b}_j \hat{\cal G}_{ji}(\varepsilon) \delta \hat{v}_i^{p}
\right)
\right\}.
\label{eqn:fij}
\end{equation}
By comparing it with Eq.~(\ref{eqn:spinf}), using Eq.~(\ref{eqn:Jij}) for $J_{ij}$, and the values of
$\delta \boldsymbol{e}_i$ and $\delta \boldsymbol{e}_j$, derived from SCLR, one can find
$[ \boldsymbol{d}_{ij}$$\times$$\boldsymbol{e}^0]$. Then, for $\boldsymbol{e}^0 = (0,0,1)$, this procedure gives us
the $x$- and $y$-projections of $\boldsymbol{d}_{ij}$. By repeating these calculations for other directions of $\boldsymbol{e}^0$,
one can find all three projections of $\boldsymbol{d}_{ij}$.

  Eq.~(\ref{eqn:fij}) was first applied in Ref.~\onlinecite{PRL96}, where
the total perturbation ($\delta \hat{v}^{p}$) was replaced by the `bare' SO interaction
($\delta \hat{v}^{ext}$, in the notations of Sec.~\ref{sec:LR}). Since the
perpendicular components of the magnetization, $\{ \delta \boldsymbol{e}_i \}$, do not contribute to $\delta \hat{v}^{ext}$,
in the latter case we need not to consider the second term in Eq.~(\ref{eqn:spinf}).
As was explained in the Introduction, this procedure is justified when
it is combined with two additional approximation for $\hat{v}$: LSDA
and ASA (that was indeed the case in Ref.~\onlinecite{PRL96}).
Then, the potential $\hat{v}$ does not depend
on the orbital degrees of freedom and the SO interaction will contribute to $\delta \hat{v}$ only
in the second order. Only in the latter case, (i) $\delta \hat{v}^{p}$ in Eq.~(\ref{eqn:fij})
can be replaced by $\delta \hat{v}^{ext}$; and (ii)
one can apply the magnetic force theorem also for the analysis of the MA energy.\cite{PRB95}
However, this procedure is no longer valid in the case of the orbital-dependent potential:
as we will see in Sec.~\ref{sec:Results},
the use of the total perturbation $\delta \hat{v}^{p}$ in Eq.~(\ref{eqn:fij}) is essential in order to reproduce parameters of DM interactions
and details of the spin canting.
Furthermore,
in the framework of the SCLR theory, the parameters of magnetocrystalline anisotropy can be derived only from the \textit{total energy},
but not from the single-particle energy. This is the reason why we do not consider here
the perturbation, caused by $\delta \hat{v}^{p}$ at both magnetic sites (similar to Ref.~\onlinecite{PRB95}):
simply, it is beyond the accuracy of the magnetic force theorem.
Nevertheless, Eq.~(\ref{eqn:dEtotal}) for the total energy suggests that such an interpretation of the MA in terms
of pairwise interactions $\tensor{\tau}_{ij}$ could be still possible by considering the mixed type of perturbation,
combining $\delta \hat{v}^{p}$ and $\delta \hat{v}^{ext}$.
We will leave this problem for future analysis.

\section{\label{sec:Results} Results}

\subsection{\label{sec:YTiO3} Spin canting in YTiO$_3$}

\begin{figure}[h!]
\begin{center}
\includegraphics[width=8cm]{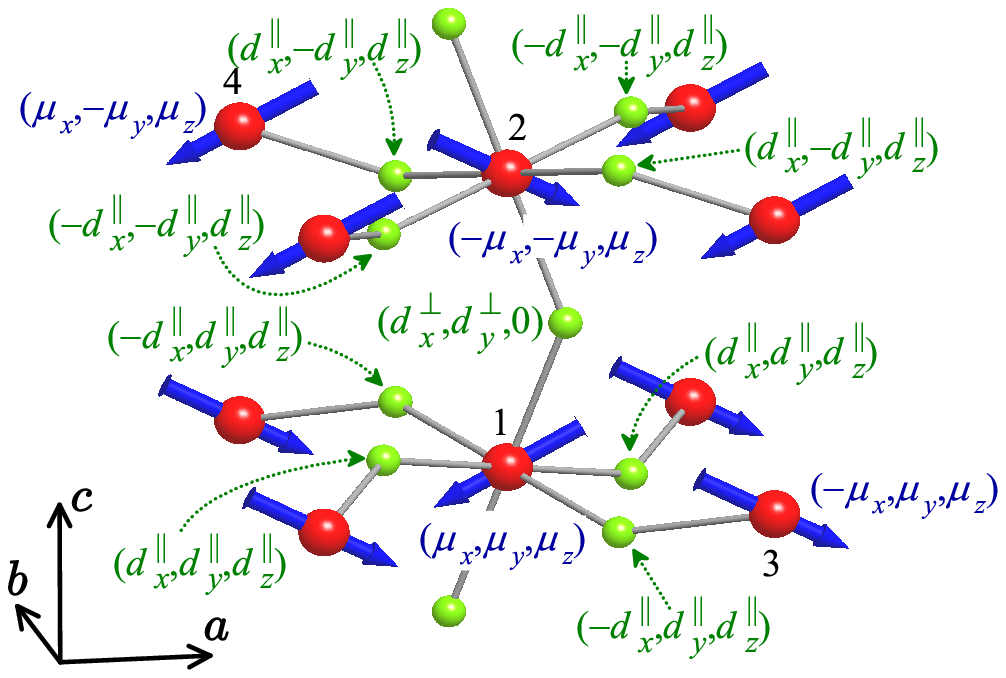}
\includegraphics[width=8cm]{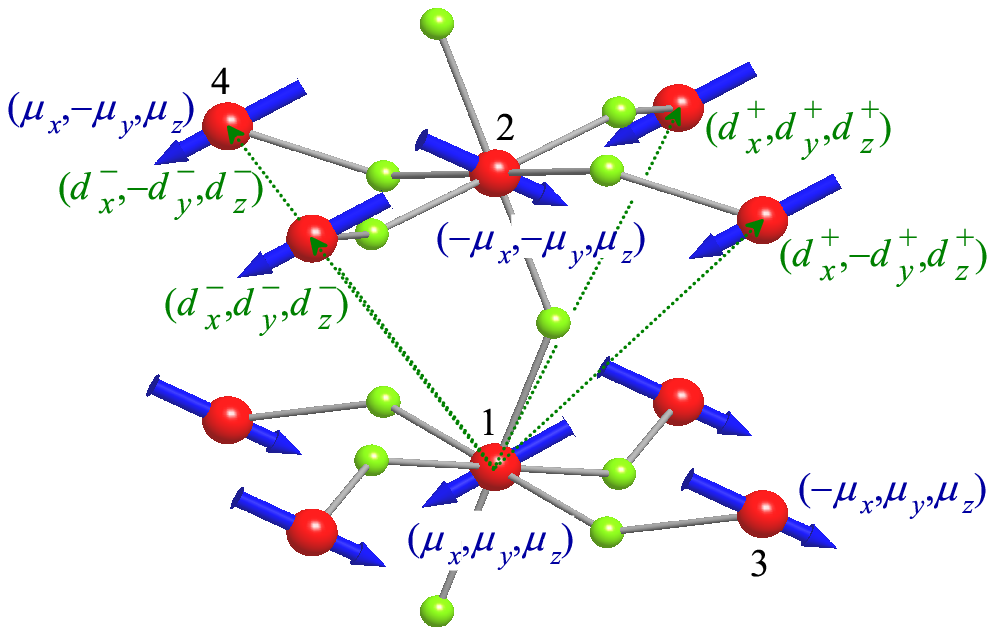}
\end{center}
\caption{\label{fig.YTO}
(Color online)
Fragment of crystal structure of YTiO$_3$ with the notation of DM interactions
and phases of magnetic moments at four Ti sites in the
primitive cell. The Ti atoms are indicated by the medium (red) spheres
and the oxygen atoms are indicated by the small (green) spheres.
Left panel explains the behavior of in-pane ($\parallel$)
and inter-plane ($\perp$) nearest-neighbor interactions.
Right panel explains the behavior of next-nearest-neighbor interactions between the planes.
In the $Pbnm$ structure, there are two types of such interactions, which are denoted as $\boldsymbol{d}^+$ and
$\boldsymbol{d}^-$.
The arrows show the directions of orbital magnetic moments, obtained in unrestricted
HF calculations for the effective model. The following conventions are used for
the notations of DM interactions: (i)
in each of the $\boldsymbol{ab}$ plane, the first site-index is always assumed to be in the center
($1$ or $2$ in the figure); (ii) all inter-plane interactions
start from the site $1$.
}
\end{figure}

  YTiO$_3$ crystallizes in the orthorhombic $Pbnm$ structure, which contains four Ti sites in the primitive cell
(see Fig.~\ref{fig.YTO}).\cite{YTiO3exp}
Without SO coupling,
YTiO$_3$ has FM ground state, which was successfully reproduced by
HF calculations for the effective model (\ref{eqn.ManyBodyH}),
constructed in the basis of Wannier functions, as explained in Sec.~\ref{sec:LEmodel}.\cite{review2008,t2g}

  The effect of the SO coupling was also studied on the level of HF calculations.\cite{review2008,t2g}.
In this section, we will be mainly interested in how well the results of
fully self-consistent non-perturbative HF calculations for YTiO$_3$ will be
reproduced by the SCLR theory.
For these purposes we start with the self-consistent
HF potential without the SO coupling, align the FM magnetization consequently along the orthorhombic
$\boldsymbol{a}$, $\boldsymbol{b}$, or $\boldsymbol{c}$ axes, switch on the SO coupling,
and again self-consistently solve the
HF equations either directly or in the framework of the SCLR theory.
In all these calculations we use the experimental structure parameters, measured at 2 K.\cite{YTiO3exp}

  The first important question is the behavior MA energy,
which specifies the direction of the FM magnetization in the ground state. In Table~\ref{tab:YTOM},
we list the total energy differences, obtained for different directions of the FM magnetization.
\begin{table}[h!]
\caption{Comparison of results, obtained in the Hartree-Fock (HF) method and
in self-consistent linear response (SCLR) theory for YTiO$_3$:
spin $(\,{\mu}^S_x,\,{\mu}^S_y,\,{\mu}^S_z)$ and orbital $(\,{\mu}^L_x,\,{\mu}^L_y,\,{\mu}^L_z)$ magnetic moments
in the ground state (in $\mu_{\rm B}$), and the total energy differences (in meV per one formula unit) between
three magnetic configurations, in which the FM magnetization was
parallel to the orthorhombic
$\boldsymbol{a}$ ($E_{||a}$), $\boldsymbol{b}$ ($E_{||b}$), or $\boldsymbol{c}$ ($E_{||c}$) axes.
The total energies in SCLR were obtained in the third and second order of the SO coupling
(the second-order results are shown in parenthesis). The third-order calculations are based on the
$2n$$+$$1$ theorem.
The phases of magnetic moments at different Ti-sites
are explained in Fig.~\ref{fig.YTO}.}
\label{tab:YTOM}
\begin{ruledtabular}
\begin{tabular}{ccccc}
 method  & $(\,{\mu}^S_x,\,{\mu}^S_y,\,{\mu}^S_z)$      & $(\,{\mu}^L_x,\,{\mu}^L_y,\,{\mu}^L_z)$  &
 $E_{||a}$$-$$E_{||c}$ & $E_{||b}$$-$$E_{||c}$ \\
\hline
  HF     & $(\,-$$0.021,\,-$$0.127,\,0.986)$            & $(\,-$$0.033,\, -$$0.001,\, -$$0.018)$  & $0.074$ & $0.067$ \\
 SCLR    & $(\,-$$0.024,\,-$$0.115,\,1.000)$            & $(\,-$$0.036,\, -$$0.010,\, -$$0.019)$  & $0.069$ ($0.072$) & $0.069$ ($0.061$) \\
\end{tabular}
\end{ruledtabular}
\end{table}
The SCLR nicely reproduces results of regular HF calculations: the total energies,
obtained using different methods, agree within 10 \%.
Generally, the use of $2n$$+$$1$ theorem provides a much agreement with the HF method for the total energies.
These results also confirm that the FM magnetization is parallel to the $\boldsymbol{c}$ axis, in agreement with the
experiment.\cite{Ulrich2002}

  Then, let us consider fine details of the magnetic structure such as the spin canting and the behavior of orbital magnetization.
If the FM magnetization is parallel to $\boldsymbol{c}$, the ground-state magnetic structure can be abbreviated G-A-F,
where G, A, and F is the type of the magnetic ordering (G-AFM, A-AFM, and FM),
formed by the $\boldsymbol{a}$, $\boldsymbol{b}$, and $\boldsymbol{c}$ projections of the magnetic moments, respectively.
This imposes some constraint on the phases of magnetic moments at different Ti sites,
which are explained in Fig.~\ref{fig.YTO}.
The values of spin and orbital magnetic moments, obtained in the HF calculations and in the framework of the SCLR theory for the G-A-F
ground state, are summarized in Table~\ref{tab:YTOM}.
Again, we note an excellent agreement between SCLR and non-perturbative HF calculations.
Nevertheless, YTiO$_3$ is somewhat special example for such a comparison: due to the $d^1$ configuration of the ion Ti$^{3+}$,
there will be no SI anisotropy term, and relative directions of the spins will be mainly controlled
by the DM interactions, which emerge in the first order of the SO coupling. This partly explains why SCLR,
which is also a first-order theory, works exceptionally well for YTiO$_3$. If the SI anisotropy is large and controls
the directions of spin magnetic moments, the agreement may not be so good. We will see it in Sec.~\ref{sec:LaMnO3}, in the
case of LaMnO$_3$.

  Finally, we discuss the origin of the spin canting and
and consider how well this canting can be reproduced by parameters of the
spin model, Eq.~(\ref{eqn:spinE}), in the frameworks of SCLR.
The parameters of isotropic and DM interactions between neighboring Ti sites are summarized in Table~\ref{tab:YTODM},
together with the DM parameters, obtained in FPA.
\begin{table}[h!]
\caption{Parameters of nearest-neighbor isotropic interactions ($J$) and DM interactions $(d_x,d_y,d_z)$ in the
$\boldsymbol{ab}$ plane of YTiO$_3$ (denoted by $\parallel$) and between neighboring planes (denoted by $\perp$).
The units are meV.
The DM interactions were computed using both
frozen potential approximation (FPA) and the self-consistent linear response (SCLR) theory.
`SE' denote results of superexchange calculations reported in Ref.~\onlinecite{NJP09}
for the same crystal structure of YTiO$_3$.
The phases of DM interactions in the orthorhombic $Pbnm$ structure
are explained in Fig.~\ref{fig.YTO}.}
\label{tab:YTODM}
\begin{ruledtabular}
\begin{tabular}{ccccc}
 method  & $J^{\parallel}$ & $(d_x^{\,\parallel},d_y^{\,\parallel},d_z^{\,\parallel})$ & $J^{\perp}$ &  $(d_x^{\perp},d_y^{\perp},d_z^{\perp})$ \\
\hline
$\begin{array}{c}
\textrm{FPA}   \\
\textrm{SCLR} \\
\end{array}$ &
$3.83$      &
$\begin{array}{c}
(\,-$$0.041,\,-$$0.031,\,-$$0.009) \\
(\,-$$0.237,\,-$$0.087,\,-$$0.043) \\
\end{array}$ &
$\phantom{-}$$0.97$ &
$\begin{array}{c}
(\,0.025,\,-$$0.011,\,0) \\
(\,0.026,\,-$$0.019,\,0) \\
\end{array}$ \\
SE & $2.90$ & $(-$$0.424,\,-$$0.367,\,-$$0.134)$ & $-$$0.18$ & $(\,0.306,\,-$$0.104,\,0)$ \\
\end{tabular}
\end{ruledtabular}
\end{table}
The phases of DM interactions, associated with different bonds, are explained in Fig.~\ref{fig.YTO}.

  Let us start with the collinear FM structure and align the spin moments parallel to the $\boldsymbol{c}$ ($z$) axis.
Then, DM interactions
give rise to other components of the spin magnetization direction, $e_x$ and $e_y$, which are parallel to
the orthorhombic axes
$\boldsymbol{a}$ and $\boldsymbol{b}$, respectively.
The corresponding energy gain (per one Ti site) is given
in the first order of $e_x$ and $e_y$
by
$\delta E_{\rm DM} = 2d_x^{\perp}e_y-(2d_y^{\perp}$$+$$4d_y^{\,\parallel})e_x$.
This spin canting acts against isotropic exchange interactions. In the second order of $e_x$ and $e_y$,
the corresponding energy loss
is given by
$\delta E_{\rm H} = J^{\perp}(e_x^2$$+$$e_y^2) + 2J^{\parallel}e_x^2$.
By minimizing $\delta E_{\rm DM}$ and $\delta E_{\rm H}$ with respect to $e_x$ and $e_y$, one finds:
$e_x = (d_y^{\perp}$$+$$2d_y^{\,\parallel})/(J^{\perp}$$+$$2J^{\parallel})$ and
$e_y = -$$d_x^{\perp}/J^{\perp}$.
Using parameters of nearest-neighbor (NN) interactions from Table~\ref{tab:YTODM}, $e_x$ and $e_y$ can be estimated
in SCLR as $-$$0.022$ and $-$$0.027$, respectively.
The value of $e_x$ is well consistent with
${\mu}^S_x/|\boldsymbol{\mu}^S| \approx -$$0.024$
obtained from the electronic model (Table~\ref{tab:YTOM}), while $e_y$ is underestimated by factor four.

  Nevertheless, much better agreement with the electronic model
can be obtained by considering next-NN magnetic interactions between the planes.
In the $Pbnm$ structure, there are two types of such interactions:
if $(R_x,R_y,R_z) = (\pm\frac{a}{2},\pm$$\frac{b}{2},\pm$$\frac{c}{2})$ are the radius-vectors,
connecting two magnetic sites, the superscripts $+$ and $-$ will denote the next-NN interactions in the bonds with $R_xR_z > 0$
and $R_xR_z < 0$, respectively (see Fig.~\ref{fig.YTO}). Then, the additional energy gain,
caused by $\boldsymbol{d}^+$ and $\boldsymbol{d}^-$, is given by $\delta E_{\rm DM}' = 4(d^+_x$$+$$d^-_x)e_y$.
In the G-A-F magnetic structure, these DM interactions will affect only the $e_y$ component of the
spin magnetization (see phases of magnetic moments and DM interactions in Fig.~\ref{fig.YTO}).
Corresponding energy loss due to isotropic interactions is given by $\delta E_{\rm H}' = 2(J^+$$+$$J^-)e_y^2$. The SCLR yields
the following parameters (in meV): $\boldsymbol{d}^+ = (\,-$$0.010,\,-$$0.002,\,0)$,
$\boldsymbol{d}^- = (\,0.040,\,0.004,\,0.014)$,
$J^+ = 0.11$, and $J^- = -$$0.09$. Therefore, it is clear that relatively large
$d^-_x$ will be responsible for additional spin canting along $y$. Indeed, by minimizing
$\delta E_{\rm DM}$, $\delta E_{\rm DM}'$, $\delta E_{\rm H}$, and $\delta E_{\rm H}'$, one finds
$e_y = -$$0.085$, which is in reasonable agreement with
${\mu}^S_y/|\boldsymbol{\mu}^S| \approx -$$0.114$, obtained in the electronic model (Table~\ref{tab:YTOM}).
Thus, parameters of spin model in the SCLR theory, well reproduce results of electronic model in the HF approximation.
FPA substantially underestimates parameters of DM interactions (see Table~\ref{tab:YTODM}) and, therefore, the spin canting.

  For the $d^1$ compounds, parameters of spin model can estimated using the theory of superexchange (SE) interactions,\cite{NJP09}
which yields somewhat different values of the parameters of isotropic and DM interactions (see Table~\ref{tab:YTODM}).
This demonstrates complexity
of the problem.
The SCLR formalism, developed in Sec.~\ref{sec:DM}, is applicable only for small deviations near the nonrelativistic
ground state. It does not work for the large spin canting. On the other hand, the theory of SE interactions is applicable
for any canting, but only in the limit of large one-site Coulomb repulsion. The SE theory also takes into account
some correlation interactions at the atomic sites, beyond the HF approximation, which additionally stabilize AFM
interactions and, therefore, enhances the spin canting away from the FM state.\cite{review2008,t2g,NJP09}

\subsection{\label{sec:LaMnO3} Weak ferromagnetism in LaMnO$_3$}

  LaMnO$_3$ crystallizes in the orthorhombic $Pbnm$ structure, similar to YTiO$_3$.
In this study we use the experimental structure parameters, reported in Ref.~\onlinecite{LaMnO3exp}.

  Without SO interaction, LaMnO$_3$ forms the collinear A-type AFM structure, which was successfully
reproduced by unrestricted HF calculations for the effective model.\cite{JPSJ} The SO interaction
results in a small canting of spins. The new magnetic ground state is of the G-A-F type, similar to
YTiO$_3$.\cite{PRL96} The main purpose of this section is to explore how well the details of this
magnetic ground state can be reproduced by the SCLR theory. Thus, we start with the collinear A-type
AFM structure, switch on the SO coupling, and compare results of straightforward HF and SCLR calculations.

  The MA energy is well reproduced by SCLR (see Table~\ref{tab:LMOM}), especially in the third-order
calculations for the SO coupling, based on the $2n$$+$$1$ theorem. These calculations also confirm that
the main A-type AFM component of the magnetization is parallel to the orthorhombic $\boldsymbol{b}$ axis,
in agreement with the experiment.\cite{Matsumoto} SCLR also nicely reproduces the values of the orbital moments,
which are in good agreement with the results of non-perturbative HF calculations.
\begin{table}[h!]
\caption{Comparison of results, obtained in the Hartree-Fock (HF) method and
in self-consistent linear response (SCLR) theory for LaMnO$_3$:
spin $(\,{\mu}^S_x,\,{\mu}^S_y,\,{\mu}^S_z)$ and orbital $(\,{\mu}^L_x,\,{\mu}^L_y,\,{\mu}^L_z)$ magnetic moments
in the ground state (in $\mu_{\rm B}$), and the total energy differences (in meV per one formula unit) between
three magnetic configurations, in which the A-type AFM magnetization was
parallel to the orthorhombic
$\boldsymbol{a}$ ($E_{||a}$), $\boldsymbol{b}$ ($E_{||b}$), or $\boldsymbol{c}$ ($E_{||c}$) axes.
The total energies in SCLR were obtained in the third and second order of the SO coupling
(the second-order results are shown in parenthesis). The third-order calculations are based on the
$2n$$+$$1$ theorem.
The phases of magnetic moments at different sites in the orthorhombic $Pbnm$ structure
are explained in Fig.~\ref{fig.YTO}.}
\label{tab:LMOM}
\begin{ruledtabular}
\begin{tabular}{ccccc}
 method  & $(\,{\mu}^S_x,\,{\mu}^S_y,\,{\mu}^S_z)$      & $(\,{\mu}^L_x,\,{\mu}^L_y,\,{\mu}^L_z)$  &
 $E_{||a}$$-$$E_{||b}$ & $E_{||c}$$-$$E_{||b}$ \\
\hline
  HF     & $(\,0.354,\,3.952,\,0.111)$            & $(\,-$$0.030,\,-$$0.057,\,-$$0.008)$  & $0.996$ & $1.133$ \\
 SCLR    & $(\,0.165,\,3.975,\,0.044)$            & $(\,-$$0.027,\,-$$0.056,\,-$$0.007)$  & $0.996$ ($0.773$) & $1.081$ ($0.941$) \\
\end{tabular}
\end{ruledtabular}
\end{table}

  Then, the NN DM interactions between the planes give rise to the weak ferromagnetism along $\boldsymbol{c}$,
while the in-plane interactions yield the G-type AFM canting parallel to $\boldsymbol{a}$.\cite{PRL96} Corresponding energy gain
is given by $\delta E_{\rm DM} = 4 d_z^{\,\parallel} e_x  + 2 d_x^{\perp} e_z$.
The contribution of next-NN interactions (see Fig.~\ref{fig.YTO}) can be evaluated as
 $\delta E_{\rm DM}' = 4 (d_x^+$$+$$d_x^-) e_z  - 4 (d_z^+$$+$$d_z^-) e_x$.
The energy loss due to isotropic interactions is
$\delta E_{\rm H} = 2 J^{\parallel} e_x^2 - J^{\perp} e_z^2$ and $\delta E_{\rm H}' = -$$2(J^+$$+$$J^-)(e_x^2$$+$$e_z^2)$,
for the NN and next-NN interactions,
respectively.
By minimizing these four terms, and using SCLR parameters of NN interactions (Table~\ref{tab:LMODM}),
together with $\boldsymbol{d}^+ = (\,0.054,\,0.005,\,0.034)$, $\boldsymbol{d}^- = (\,0.022,\,0.008,\,0.042)$,
$J^+ = -$$1.32$, and $J^- = -$$1.16$ for the next-NN interactions (in meV), one finds $e_x = 0.064$ and
$e_z = 0.016$. These values are in good agreement with results of SCLR calculations for
$\mu^S_x/|\boldsymbol{\mu}^S| \approx 0.041$ and $\mu^S_y/|\boldsymbol{\mu}^S| \approx 0.011$
(Table~\ref{tab:LMOM}). Thus, the spin canting in SCLR is nicely explained by the competition of isotropic
and DM interactions with the parameters derived from the magnetic force theorem.
\begin{table}[h!]
\caption{Parameters of nearest-neighbor isotropic interactions ($J$) and DM interactions $(d_x,d_y,d_z)$ in the
$\boldsymbol{ab}$ plane of LaMnO$_3$ (denoted by $\parallel$) and between the neighboring planes (denoted by $\perp$).
The units are meV.
The DM interactions were computed using both
frozen potential approximation (FPA) and the self-consistent linear response (SCLR) theory.
The phases of DM interactions in the orthorhombic $Pbnm$ structure
are explained in Fig.~\ref{fig.YTO}.}
\label{tab:LMODM}
\begin{ruledtabular}
\begin{tabular}{ccccc}
 method  & $J^{\parallel}$ & $(d_x^{\,\parallel},d_y^{\,\parallel},d_z^{\,\parallel})$ & $J^{\perp}$ &  $(d_x^{\perp},d_y^{\perp},d_z^{\perp})$ \\
\hline
$\begin{array}{c}
\textrm{FPA}   \\
\textrm{SCLR} \\
\end{array}$ &
$3.86$      &
$\begin{array}{c}
(\,-$$0.234,\,0.254,\,-$$0.250) \\
(\,-$$0.388,\,0.384,\,-$$0.328) \\
\end{array}$ &
$-$$4.47$ &
$\begin{array}{c}
(\,-$$0.070,\,0.159,\,0) \\
(\,-$$0.302,\,0.494,\,0) \\
\end{array}$ \\
\end{tabular}
\end{ruledtabular}
\end{table}

  Nevertheless, the agreement between HF and SCLR calculations for spin magnetic moments, $\boldsymbol{\mu}^S$,
is not so good as in YTiO$_3$. The reason is the SI anisotropy, which also controls the directions of local
magnetic moments in the case of LaMnO$_3$: since this is the second-order effect of the SO coupling,
it is not captured by SCLR.

  FPA underestimates the DM interactions (Table~\ref{tab:LMODM}) and, therefore, the spin canting,
even in comparison with SCLR.

  The NN DM interactions in LaMnO$_3$ were evaluated in Ref.~\onlinecite{PRL96}, by using FPA and LSDA
for the electronic structure calculations. These calculations yielded the following parameters
(apart from the phases, which depend on the choice of the origin in the lattice):
$\boldsymbol{d}^{\parallel} = (\,-$$0.435,\,0.326,\,-$$0.530)$ meV and
$\boldsymbol{d}^{\perp} = (\,-$$0.435,\,0.707,\,0)$ meV, which are in reasonable agreement
with results of the SCLR calculations in the present work (see Table~\ref{tab:LMODM}). Nevertheless,
this agreement is somewhat fortuitous: LSDA itself does not include the orbital polarization effects and,
in this sense, is a poor approximation for the analysis of DM interactions. On the other hand, it justifies the
use of FPA. This is the main reason why the combination of these two approximation provides a reasonable
estimate for the parameters of DM interactions.

\subsection{\label{sec:BiFeO3} Spiral magnetic ordering in BiFeO$_3$}

  Below $1100$ K, BiFeO$_3$ crystallizes in the noncentrosymmetric rhombohedral $R3c$ structure,
which allows for the ferroelectricity. Around $650$ K, it undergoes the magnetic transition to the
G-type AFM phase. The most interesting aspects from the viewpoint of magnetism are related to the
emergence of DM interactions in the noncentrosymmetric structure, which give rise to such phenomena as
the spin-spiral modulation of the collinear G-type AFM order with the period of
$620$~\AA (Ref.~\onlinecite{Kadomtseva,Sosnowska,BiFeO3lattice,BiFeO3Jeong,BiFeO3Matsuda}) and
the local weak ferromagnetism (Refs.~\onlinecite{Kadomtseva,EdererSpaldin,Ramazanoglu}).

  We use the experimental rhombohedral crystal structure with the
lattice parameters $a_{\rm H}=5.581$ \AA~and $c_{\rm H}=13.876$ \AA, reported in Ref.~\onlinecite{BiFeO3lattice} (in the hexagonal settings).
The rhombohedral lattice translations are given by $(a,0,c)$, $(-$$a/2,\sqrt{3}a/2,c)$, and $(-$$a/2,-$$\sqrt{3}a/2,c)$, where
$a = a_{\rm H}/\sqrt{3}$ and $c = c_{\rm H}/3$. In the following, we will operate with the parameters $a$ and $c$.

  The space group $R3c$
can be generated by two symmetry operations:
$\hat{C}^{3+}_z$, that is the clockwise threefold rotation around the $z$ axis, and
$\{ \hat{m}_y | (0,0,3c/2) \}$,
that is the mirror reflection $y \rightarrow -$$y$ combined with the translation by $(0,0,3c/2)$.
Thus, all NN Fe-Fe bonds can be obtained from a single bond
(say, $0$-$1$ in Fig.~\ref{fig.BiFeO3}) by applying the following symmetry operations (apart from primitive translations):
\begin{figure}[h!]
\begin{center}
\includegraphics[width=10cm]{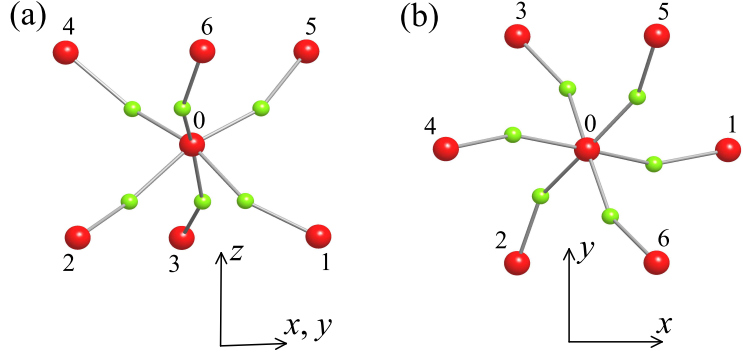}
\end{center}
\caption{\label{fig.BiFeO3}
(Color online)
Fragment of the $R3c$ structure of BiFeO$_3$: side view (a) and top view (b).
The Fe atoms are indicated by the medium (red) spheres
and the oxygen atoms are indicated by the small (green) spheres.}
\end{figure}
$\boldsymbol{R}_{02} = \hat{C}^{3+}_z \boldsymbol{R}_{01}$, $\boldsymbol{R}_{03} = \hat{C}^{3-}_z \boldsymbol{R}_{01}$,
$\boldsymbol{R}_{04} = -$$\hat{m}_y \boldsymbol{R}_{01}$, $\boldsymbol{R}_{05} = -$$\hat{C}^{3+}_z \hat{m}_y \boldsymbol{R}_{01}$,
and $\boldsymbol{R}_{06} = -$$\hat{C}^{3-}_z \hat{m}_y \boldsymbol{R}_{01}$,
where $\boldsymbol{R}_{ij}$ is the radius-vector connecting the site $i$ with the site $j$, and
$\hat{C}^{3-}_z \equiv (\hat{C}^{3+}_z)^2$ is the counterclockwise rotation around $z$. Therefore, all parameters of isotropic ($J_{0i}$) and
DM ($\boldsymbol{d}_{0i}$) NN interactions can be obtained from the ones in an arbitrarily taken bond $0$-$1$,
for which we adopt the shorthand notations: $\boldsymbol{R}_{01} \equiv \boldsymbol{R}$ and $\boldsymbol{d}_{01} \equiv \boldsymbol{d}$.
This means that all scalar parameters will be identical, $J_{0i} \equiv J$, while the vectors $\boldsymbol{d}_{0i}$ satisfy
the following properties:
$\boldsymbol{d}_{02} = \hat{C}^{3+}_z \boldsymbol{d}$, $\boldsymbol{d}_{03} = \hat{C}^{3-}_z \boldsymbol{d}$,
$\boldsymbol{d}_{04} = -$$\hat{m}_y \boldsymbol{d}$, $\boldsymbol{d}_{05} = -$$\hat{C}^{3+}_z \hat{m}_y \boldsymbol{d}$,
and $\boldsymbol{d}_{06} = -$$\hat{C}^{3-}_z \hat{m}_y \boldsymbol{d}$.
Moreover, since $\boldsymbol{d}$
is an \textit{axial} vector, the operation $\hat{m}_y \boldsymbol{d}$ actually reads as
$\hat{C}^2_y \boldsymbol{d}$, where $\hat{C}^2_y$ is the twofold rotation around $y$.
Finally, there is no symmetry restriction
on the form of $\boldsymbol{d}$, which is characterized by three independent parameters:
$\boldsymbol{d} \equiv (d_x,d_y,d_z)$. Amongst them,
$d_x$ and $d_y$
are responsible for the formation of the spiral magnetic ordering (see Appendix), while
$d_z$ gives rise to the weak ferromagnetism in the $xy$-plane.\cite{Kadomtseva,Ramazanoglu}

  The numerical calculations in FAP and
SCLR yield
$\boldsymbol{d} =$  $(\,0.145, \, -$$0.418, \, 0.177)$ and
$(\,0.494, \,-$$1.450, \,0.330)$, respectively (in meV). Thus, similar to previous examples,
all DM interactions are strongly enhanced in SCLR.
Therefore, it is important to check whether these values are consistent with the experimental period of the spin spiral
and the weak FM moment in
BiFeO$_3$.

  First, let us evaluate other parameters, which are necessary for our analysis. Using the theory of infinitesimal
spin rotations near the G-type AFM state,\cite{review2008,JHeisenberg}
we obtain
$J$$=$ $-$$37.27$ meV
for the NN interactions and $J'$$=$ $-$$1.60$ meV
for the six next-NN interactions in the $xy$-plane. These values are consistent
with available experimental data.\cite{BiFeO3Jeong,BiFeO3Matsuda,remark1} Furthermore,
using these parameters, one can evaluate the theoretical N\'eel temperature. For these
purposes we use Tyablikov's random-phase approximation (Ref.~\onlinecite{spinRPA}), which yields
$T_{\rm N} \sim 785$ K, being in reasonable agreement with the experimental value of $ 650$ K.\cite{BiFeO3Jeong}

  The weak ferromagnetism is expected in the primitive cell of BiFeO$_3$, containing two formula units, when
spins lie in the $xy$-plane. In Table~\ref{tab:BFOWF} we summarize results of
self-consistent non-perturbative HF calculations and SCLR method
for the in-plane ($\boldsymbol{\mu}^S \perp z$) and out-of-plane ($\boldsymbol{\mu}^S || z$) configurations
of spins. As expected for the $d^5$ configuration of the ions Fe$^{3+}$, the orbital magnetization is small
and can be neglected in the present analysis.
\begin{table}[h!]
\caption{Vectors of spin magnetic moments (in $\mu_{\rm B}$) and corresponding total energy difference (in meV per one formula unit)
for the in-plane ($\perp$$z$) and out-of-plane ($||$$z$)
configurations of spins,
as obtained in the self-consistent non-perturbative Hartree-Fock (HF) calculations and in the self-consistent linear response (SCLR) theory
for BiFeO$_3$. The total energies in SCLR were obtained in the third and second order of the SO coupling
(the second-order results are shown in parenthesis). The third-order calculations are based on the $2n$$+1$ theorem.
In the notations $\pm$ and $\mp$, the upper sign
corresponds to the central site `0' in Fig.~\ref{fig.BiFeO3}, while the lower sign corresponds
to its neighboring sites, belonging to another magnetic sublattice
in the G-type AFM structure.}
\label{tab:BFOWF}
\begin{ruledtabular}
\begin{tabular}{cccc}
 method  & $\boldsymbol{\mu}^S || z$ & $\boldsymbol{\mu}^S$$\perp$$z$ & $E_{||z}$$-$$E_{\perp z}$ \\
\hline
  HF     & $(\,0,\,0,\,\pm$$4.890)$            & $(\,\mp$$4.890,\,0.042,\,0)$             & $0.135$             \\
 SCLR    & $(\,0,\,0,\,\pm$$4.899)$            & $(\,\mp$$4.899,\,0.041,\,0)$             & $0.135$ ($0.119$)   \\
\end{tabular}
\end{ruledtabular}
\end{table}
Both methods produce very similar values of spin magnetic moments, including the weak FM component along $y$.
Moreover, the FM canting can be reproduced, even quantitatively, using parameters of isotropic and DM interactions,
obtained in the SCLR scheme. Indeed, the canting of spins will lead to the
energy gain $\delta E_{\rm DM} = -$$6d_ze_y$ (per one formula unit), associated with DM interactions, and the the energy loss
$\delta E_{\rm H} = -$$3Je_y^2$, associated with isotropic interactions.
By minimizing these two contribution with respect to $e_y$, it is easy to find that $e_y=-d_z/J$, which yields
$e_y \approx 0.0089$, being in excellent agreement with $e_y = \mu^S_y/|\boldsymbol{\mu}^S| \approx 0.0086$,
obtained using results of HF calculations for $\boldsymbol{\mu}^S$ (Table~\ref{tab:BFOWF}). Thus, the parameters of spin Hamiltonian, based on
SCLR, nicely
reproduce results of electronic model. On the contrary, FPA underestimates $d_z$ and, therefore,
the spin canting. Note that the ratio $|\mu^S_y/\mu^S_x|$ corresponds to the rotation angle of about $0.5^\circ$,
which is consistent with the experimental estimate ($\sim 1^\circ$).\cite{Ramazanoglu}

  Parameters of the SI anisotropy can be extracted from the total energy difference $E_{||z}$$-$$E_{\perp z}$,
reported in Table~\ref{tab:BFOWF}. Since we are interested in the uniaxial anisotropy, it is more appropriate to use
the second order contribution $E_{||z}$$-$$E_{\perp z} = 0.119$ meV. Furthermore, this energy
contains two contributions. One is the proper SI anisotropy energy, $E_{\rm SI}$, and another one is the energy of DM interactions,
which contribute to $E_{\perp z}$ but not to $E_{||z}$, where the spins are collinear.
The contribution of DM interactions can be easily estimated using the above expression,
$-$$6d_ze_y$, which yields $0.018$ meV. Thus, $E_{\rm SI}$ is about $0.1$ meV, which is about two times larger than the
experimental value.\cite{BiFeO3Matsuda,remark1} However, it should be noted that
there is some ambiguity in separating the contributions of SI anisotropy and DM interactions
in the experiment.\cite{BiFeO3Matsuda}
Finally, parameters of the SI anisotropy tensor can be obtained from $E_{\rm SI}$ as
$\tau_{zz} = -$$2\tau_{xx} = -$$2\tau_{yy} = \frac{2}{3} E_{\rm SI}$,
which yields $\tau_{xx} = \tau_{yy} = -$$0.034$ meV and $\tau_{zz} = 0.068$ meV. Thus, $E_{\rm SI}$ favors the
in-plane configuration of spins.

  Thus, we note that the parameters of single-ion anisotropy
are about order of magnitude smaller than $d_x$ and $d_y$. In such a situation, the incommensurate spiral magnetic ordering
in BiFeO$_3$ arises mainly from the competition of DM and isotropic exchange interactions, as explained in the Appendix.
The propagation vector, corresponding to the minimum of energy, is $\boldsymbol{q} = (\delta q_x,0,2\pi/c)$, where
\begin{equation}
\delta q_x = \frac{d_y}{a(J-3J')}.
\label{eqn:BFOqx}
\end{equation}
Thus, $d_y$ can be regarded as an effective DM interaction,
responsible for spiral magnetic ordering. This interaction has been also measured experimentally.\cite{BiFeO3Jeong,BiFeO3Matsuda,remark1}
Our value $|d_y| = 1.450$ meV, obtained in SCLR, is somewhat larger than the experimental one,\cite{remark1}
which results in smaller periodicity $L$ of the spin-spiral structure.
Indeed, $L$ should be found from the condition
$aLq_x = 2 \pi$, which yields $L \approx 140$ and, therefore, $La \approx 450$ \AA, while
the experimental value is about $620$ \AA.
On the other hand, the value $|d_y| = 0.205$ meV, obtained in FPA, yields $La \approx 1570$ \AA,
which exceeds the experimental periodicity by more than factor two.

  There maybe several reasons why our theoretical value of $L$ in SCLR is somewhat smaller than the experimental one.
Of course, the DM interaction is a delicate quantity, which may depend on numerical factors and approximations,
underlying the construction and solution of the model Hamiltonian (\ref{eqn.ManyBodyH}). Nevertheless, there might be also
a physical reason. On the experimental side, it was emphasized that the contributions of DM interactions and
SI anisotropy cannot be easily separated.\cite{BiFeO3Matsuda}
However, if the magnetic structure was indeed the spin spiral, its periodicity should not depend on the SI anisotropy
(see Appendix and Ref.~\onlinecite{Sosnowska}). Thus,
in order to contribute to the periodicity
$L$, the SI anisotropy should deform the spin-spiral alignment and produce some inhomogeneity in the distribution
of spins. Strictly speaking, the magnetic structure in this case will not longer be the spin spiral and its periodicity
is no longer described by the simplified expression (\ref{eqn:BFOqx}).
In fact, details of the magnetic ordering in BiFeO$_3$ continue to be disputed and not completely resolved issue.\cite{Przenioslo}
The deformation of spiral magnetic ordering by the SI anisotropy is also well known for the rare-earth compounds.\cite{REbunching}

\subsection{\label{sec:BiMnO3} BiMnO$_3$: ferroelectricity and ferromagnetism, induced by antiferromagnetic inversion symmetry breaking}

  BiMnO$_3$ is one of the most important compounds in the field multiferroics, and also one of the most controversial ones.
In some sense, the new wave research activity on multiferroics was strongly influenced by the study on BiMnO$_3$,\cite{SeshadriHill}
where the ferroelectricity was believed to coexist with the ferromagnetism because of two independent mechanisms:
the lone pair effect, which
leads to the noncentrosymmetric atomic displacements, and a peculiar orbital ordering, which
gives rise to the ferromagnetism. However, this point of view was questioned by subsequent experimental studies
(Ref.~\onlinecite{belik_07}) and electronic structure calculations (Ref.~\onlinecite{spaldin_07}), which suggests
that BiMnO$_3$ should crystallize in the centrosymmetric (and, therefore, non-ferroelectric) $C2/c$ structure.
A ``compromised'' point of view was proposed in Ref.~\onlinecite{BiMnO3}, where it was argued that BiMnO$_3$
could be an \textit{improper} multiferroic, where the inversion symmetry is broken by some hidden
(and not yet experimentally observed)
AFM order. In this section, we will illustrate how the proposed SCLR method can be used for microscopic analysis of
multiferroic coupling, using BiMnO$_3$ as an example. Particularly, we will show
that the magnetic inversion symmetry breaking is not only responsible for the ferroelectricity, but
can also induce the DM interactions across the inversion centers. In BiMnO$_3$, these DM interactions are
responsible for the FM magnetization.

  First, we will briefly remind the reader the main results of Ref.~\onlinecite{BiMnO3}.

\begin{figure}[h!]
\begin{center}
\includegraphics[width=15cm]{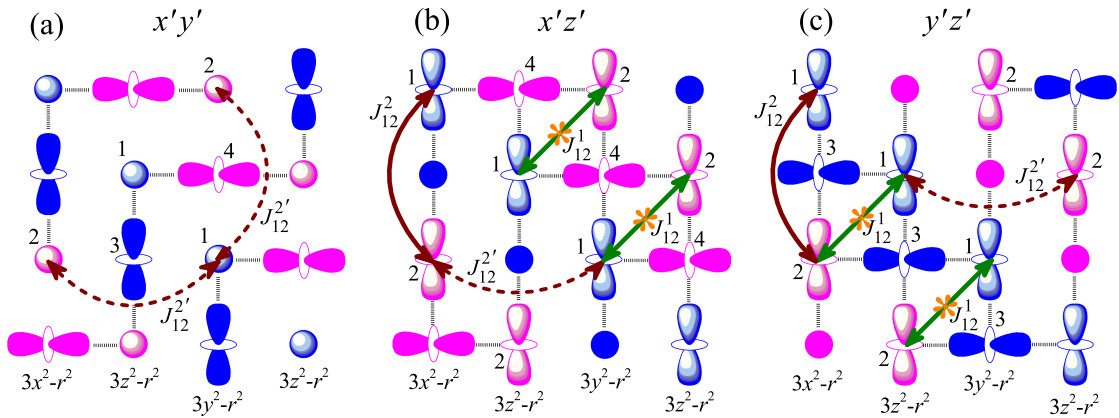}
\end{center}
\caption{\label{fig.BiMnO3OO}
(Color online)
Schematic view on the orbital ordering
and corresponding interatomic magnetic interactions in the
pseudocubic $x'y'$, $x'z'$, and $y'z'$ planes of BiMnO$_3$.
The inversion centers are marked by $*$.
In the $C2/c$ phase of BiMnO$_3$, there are two groups of Mn atoms,
which are denoted as $(1,2)$ and $(3,4)$. The inversion operation
transforms the site $1$ to the site $2$ (and vice versa), and the
sites $3$ and $4$ to the equivalent sites of the same type
($3$ and $4$, respectively).
The nearest-neighbor ferromagnetic
interactions are denoted by hatched bonds.
The leading ``super-superexchange'' interactions between atoms $1$ and $2$
of the first and second coordination sphere are denoted as $J_{12}^1$ and $J_{12}^2$,
respectively. Another (weak) super-superexchange interactions is
denoted as $J_{12}^{2'}$.
}
\end{figure}

\begin{figure}[h!]
\begin{center}
\includegraphics[width=10cm]{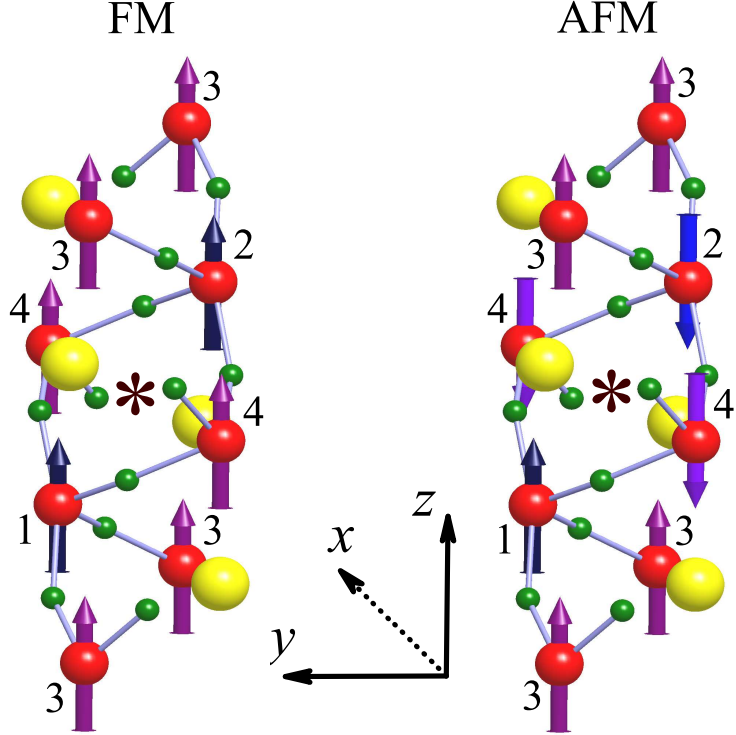}
\end{center}
\caption{\label{fig.BiMnO3FM_AFM}
(Color online)
Fragment of the crystal
structure of BiMnO$_3$. The Bi
atoms are indicated by the big light gray (yellow) spheres, the Mn
atoms are indicated by the medium gray (red) spheres, and the
oxygen atoms are indicated by the small gray (green) spheres. The
directions of magnetic moments in the ferromagnetic (FM)
and $\uparrow \downarrow \uparrow \downarrow$ antiferromagnetic (AFM)
states without spin-orbit coupling are shown by arrows. The
inversion center is marked by the symbol $*$. The central part of
the figure explains the orientation of the Cartesian coordinate frame.
}
\end{figure}

  \textit{Details of the crystal structure}.
The primitive cell of BiMnO$_3$ in the $C2/c$ phase contains four Mn sites, which form two inequivalent groups: $(1,2)$ and $(3,4)$
(see Figs.~\ref{fig.BiMnO3OO} and \ref{fig.BiMnO3FM_AFM} for the notations).
For understanding the multiferroic properties, it is important that the the spatial inversion
transforms the
sites $1$ and $2$ to each other, and
the sites $3$ (or $4$) to themselves (apart from the translation).

  \textit{Orbital ordering and magnetic interactions}.
The structure of isotropic exchange interactions in BiMnO$_3$ is closely related to the
alternation of occupied $e_g$ orbitals (or the orbital ordering) in the pseudocubic
planes $x'y'$, $x'z'$, and $y'z'$,
which is schematically explained in Fig.~\ref{fig.BiMnO3OO}.
The single $e_g$ electron occupies the $3z^2$$-$$r^2$ orbitals at the sites $1$ and $2$, and
the $3y^2$$-$$r^2$ and $3x^2$$-$$r^2$ orbitals at the sites $3$ and $4$, respectively.
Besides NN FM interactions, which
take place between sites with the
nearly orthogonal orbitals (for instance, $3x^2$$-$$r^2$ and $3z^2$$-$$r^2$
orbitals in the $x'$ direction), there are several long-range AFM interactions.
If the NN interactions are governed by the regular superexchange processes, according to
Goodenough-Kanamori rules,\cite{GK} the long-range interactions are caused by super-superexchange processes,
which are mediated by the states of intermediate Mn sites. There are two relatively strong long-range
interactions between sites $1$ and $2$, operating in the planes $x'z'$ and $y'z'$: $J_{12}^1$,
operating across the inversion center, and $J_{12}^2$, operating in the chains parallel to $z'$.
Another interaction $J_{12}^{2'}$ in the chains parallel to the $x'$ and $y'$ axes is considerably weaker,
due to weaker overlap of occupied $3z^2$$-$$r^2$ orbitals of the sites $1$ and $2$ in these two directions.
The values of $J_{12}^1$ and $J_{12}^2$ are listed in Table~\ref{tab:BMO}.
\begin{table}[h!]
\caption{Isotropic ($J_{12}^k$) and Dzyaloshinskii-Moriya ($\boldsymbol{d}_{12}^k$)
interactions between atoms of the magnetic sublattices
$1$ and $2$, calculated in the $\uparrow \downarrow \uparrow \downarrow$
antiferromagnetic noncentrosymmetric phase of BiMnO$_3$ (in meV). The
structure of isotropic exchange interactions is explained in Fig.~\ref{fig.BiMnO3OO}, where the leading interactions between
atoms of the first ($k$$=$$1$) and second ($k$$=$$2$)
coordination sphere are denoted as $J_{12}^1$ and $J_{12}^2$, respectively. $J_{12}^1$ operates across the
inversion centers, while $J_{12}^2$ operates in the chains parallel to the pseudocubic axis $z'$.}
\label{tab:BMO}
\begin{ruledtabular}
\begin{tabular}{ccc}
 $k$  & $J$ & $\boldsymbol{d}$ \\
\hline
  $1$     & $-$$1.28$            & $(\,-$$0.311,\,\phantom{-}$$0.040,\,0.122)$               \\
  $2$     & $-$$3.03$            & $(\,-$$0.689,\,-$$0.007,\,0.328)$               \\
\end{tabular}
\end{ruledtabular}
\end{table}
Both parameters are antiferromagnetic, that is expected for interactions between sites with the same type
of occupied orbitals.\cite{GK}

  Thus, the NN interactions alone will favor the FM coupling in the bonds
$1$-$3$, $1$-$4$, $2$-$3$, and $2$-$4$ (see Fig.~\ref{fig.BiMnO3OO}), that would lead to the
formation of the FM structure (Fig.~\ref{fig.BiMnO3FM_AFM}). On the other hands, the long-range
interactions would favor AFM coupling between sites $1$ and $2$.
Furthermore, the long-range interactions between sites $3$ and $4$
are also weakly antiferromagnetic. Therefore, the long-range interactions, if considered alone, would stabilize the AFM
$\uparrow \downarrow \uparrow \downarrow$ spin structure (Fig.~\ref{fig.BiMnO3FM_AFM}), where the arrows indicate
the relative directions of spins at the sites $1$, $2$, $3$, and $4$. This structure is equivalent to the
$\uparrow \downarrow \downarrow \uparrow$ structure, considered in Ref.~\onlinecite{BiMnO3}.

  At this point, it is instructive to make some analogy with orthorhombic manganites,
which are more studied experimentally.\cite{MF_review}
The existence of long-range AFM interactions in orthorhombic manganites, which are responsible for
the formation of complex (and, sometimes, noncentrosymmetric) magnetic structures, is also related to the orbital ordering.\cite{JPSJ,FE_model}
Moreover, the basic mechanisms, underlying the behavior of interatomic magnetic interactions,
are very similar in orthorhombic manganites and monoclinic BiMnO$_3$.
The main difference is the orbital ordering pattern, which leads to different patterns of interatomic magnetic
interactions and, therefore, the types of magnetic structures, realized in the ground state.

  \textit{Origin of magnetic inversion symmetry breaking}.
The $\uparrow \downarrow \uparrow \downarrow$ spin structure breaks the inversion symmetry. The reason is the following:
Since the sites $1$ and $2$ transform to each other by the inversion operation ($\hat{I}$)
(see Fig.~\ref{fig.BiMnO3OO}), the AFM alignment between them
requires that $\hat{I}$ should be combined with the time reversal $\hat{T}$. On the other hand, $\hat{I}$
transforms the sites $3$ (or $4$) to themselves.
From this point of view, $\hat{I}$ should enter the magnetic space group as it is
(i.e., without $\hat{T}$). The enforcement of the $\hat{I}\hat{T}$ symmetry would make the sites $3$ and $4$
nonmagnetic, that is energetically unfavorable and would lead to gigantic loss of Hund's energy
(of the order of $\frac{1}{4} J_{\rm H}|\boldsymbol{\mu}^S|^2$ per Mn site, where
$J_{\rm H} \sim 0.9$ eV is the intraatomic exchange coupling and $|\boldsymbol{\mu}^S| \sim 4 \mu_{\rm B}$ is the
spin magnetic moment of the ion Mn$^{3+}$).\cite{BiMnO3}
Therefore, the only possibility to resolve this contradiction is to break the inversion symmetry.

  \textit{Origin of ferromagnetic spin canting}. The most interesting aspect of
the magnetic symmetry breaking in BiMnO$_3$ is that this material does not
only become FE, but can also carry a net magnetic moment in the ground state
after including the SO coupling.\cite{BiMnO3}
Such a combination of ferroelectricity and \textit{ferromagnetism} is indeed very rare. Therefore, this
behavior can be very important, also from the practical point of view. Then, what is the microscopic
origin of the FM spin canting in the $\uparrow \downarrow \uparrow \downarrow$ spin structure?
Note that, in the $C2/c$ phase of BiMnO$_3$, the sites $1$ and $2$ are connected by the spatial inversion
(see Figs.~\ref{fig.BiMnO3OO} and \ref{fig.BiMnO3FM_AFM}). Therefore, from the viewpoint
of the crystal structure itself, there should be no DM interactions between these two types of sites.\cite{DM}
Nevertheless, the magnetic inversion symmetry breaking produces some changes in the electronic structure,
which may give rise to the finite DM coupling.

  In this section we estimate estimate this effect and calculate the DM interactions
between sites $1$ and $2$ in the $\uparrow \downarrow \uparrow \downarrow$ AFM state, using results of the
SCLR theory, as explained in Sec.~\ref{sec:DM}. The obtained parameters are listed in Table~\ref{tab:BMO}.
One can see that these interactions are sufficiently strong. Moreover, there is a clear correlation between the strength of
isotropic and DM interactions, and the interactions in the chains ($J_{12}^2$ and $\boldsymbol{d}_{12}^2$)
are generally stronger than the ones operating across the inversion centers ($J_{12}^1$ and $\boldsymbol{d}_{12}^1$).
As a test, we have performed similar calculations in the FM phase, which respects the inversion symmetry, and found
that all $\boldsymbol{d}_{12}$ are identically equal to zero. Thus, finite interactions $\boldsymbol{d}_{12}$
in the $\uparrow \downarrow \uparrow \downarrow$ phase
are solely induced by the magnetic inversion symmetry breaking.

  Then, we can readily estimate the spin canting, caused by the competition of $J_{12}$ and $\boldsymbol{d}_{12}$
in the $\uparrow \downarrow \uparrow \downarrow$ phase. As we will see in a moment,
the magnetocrystalline anisotropy favors the configuration where all spins lie in the $xz$ plane. Moreover, the sites
$1$ and $2$ are connected by a glide reflection, which transforms $y$ to $-$$y$.\cite{BiMnO3} Therefore,
if $(e_x,e_y,e_z)$ is
the direction of spins at the sites $1$, the corresponding to it direction at the site $2$ will be
$(-$$e_x,e_y,-$$e_z)$ (note that $\boldsymbol{e}$ is an \textit{axial} vector).
Thus, from the viewpoint of symmetry, the $y$ component of spins should be coupled ferromagnetically.
The energy gain due to the FM
spin canting along $y$ is given by $\delta E_{\rm DM} = 2(d_ze_x - d_xe_z) e_y$, where
$\boldsymbol{d} \equiv (d_x,d_y,d_z) = \boldsymbol{d}_{12}^1 + \boldsymbol{d}_{12}^2$, and the energy loss due to
isotropic exchange interactions is $\delta E_{\rm H} = - J e_y^2$, where $J = J_{12}^1 + J_{12}^2$. Thus, the FM canting of spins
in the equilibrium can be estimated as $e_y = (d_ze_x - d_xe_z)/J$. It depends on the orientation of spins in the $xz$ plane,
which is controlled by the MA energy.

\begin{figure}[h!]
\begin{center}
\includegraphics[width=12cm]{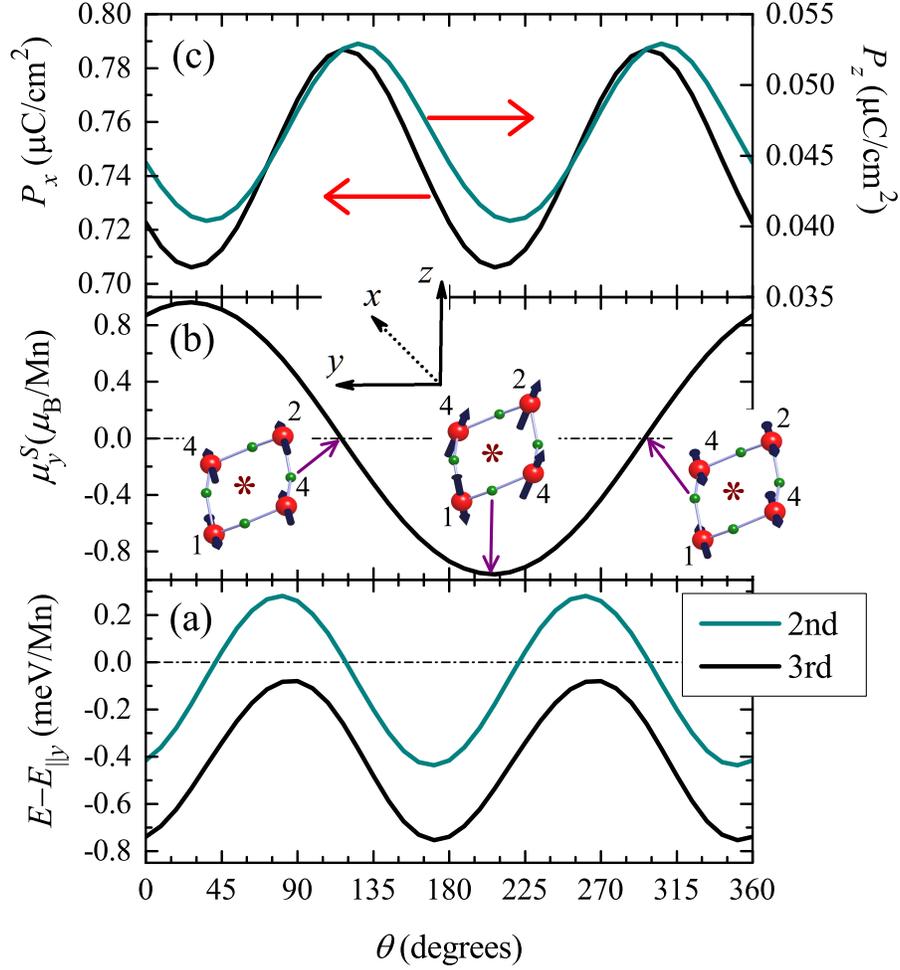}
\end{center}
\caption{\label{fig.BiMnO3summary}
(Color online)
Summary of SCLR calculations for the $\uparrow \downarrow \uparrow \downarrow$ antiferromagnetic
phase of BiMnO$_3$, where the spin magnetization is rotated in the $xz$ plane, and $\theta$ is the
polar angle, characterizing the direction of the magnetization). (a) Total energy relative
to the magnetization direction parallel to the $y$ axes in the 2nd and 3rd order of the spin-orbit
coupling (the 3rd order calculations are based on the $2n$$+$$1$ theorem). (b) Net magnetic moment
parallel to the $y$ axis. The insets show the orientation of the coordinate frame and
corresponding fragments of the crystal structure
with the directions of spin magnetic moments at the sites $1$, $2$, and $4$, around the inversion
center $*$ for different $\theta$ (see Figs.~\ref{fig.BiMnO3OO} and \ref{fig.BiMnO3FM_AFM}
for the notations of atomic sites and other details of the crystal structure).
(c) Behavior of $x$ and $z$ components of the electronic polarization.
}
\end{figure}

  The numerical details of the spin canting will depend on other interactions. Nevertheless, the above example
nicely illustrate the main idea of the SCLR calculations, which we discuss below (see Fig.~\ref{fig.BiMnO3summary}).
In this calculations with start with the AFM $\uparrow \downarrow \uparrow \downarrow$ configuration
and rotate the spins in the
$xz$ plane. This rotation is characterized by the polar angle $\theta$, such that the direction of spin
at the sites $1$ and $3$ are $\boldsymbol{e}_{1,3} = (\sin \theta, 0, \cos \theta)$ and the ones at the sites
$2$ and $4$ are $\boldsymbol{e}_{2,4} = -$$\boldsymbol{e}_{1,3}$.
Then, we switch on the SO coupling and calculate the net magnetic moment and the FE polarization,
using the Berry-phase theory,\cite{FE_theory} which was adopted for the effective Hubbard-type model
in the HF approximation.\cite{FE_model} Due to the glide reflection, $y \rightarrow -$$y$, which imposes some
symmetry constraints on the MA energy, the spin magnetization can either lie in the
$xz$ plane or be perpendicular to this plane (i.e., parallel to the $y$ axis). The in-plane configuration has lower
energy for all $\theta$, that is clearly seen in the calculations based on the $2n$$+$$1$ theorem [see Fig.~\ref{fig.BiMnO3summary}(a)].
Then, the DM interactions lead to the FM spin canting along $y$ [see Fig.~\ref{fig.BiMnO3summary}(b)].
As was discussed above,
the magnitude of this canting depends on the direction of spins in the $xz$ plane.

  The magnetic inversion symmetry breaking
gives rise to the FE activity. Again, due to the glide reflection, $y \rightarrow -$$y$, the
FE polarization lies in the $xz$ plane. Moreover, the $z$ component of the FE polarization
is substantially smaller than the $x$ one [see Fig.~\ref{fig.BiMnO3summary}(c)], being consistent with the
previous finding (Ref.~\onlinecite{BiMnO3}). Very importantly, the value of the FE polarization
``anticorrelates'' with that of the FM magnetization, $\mu_y^S$: the larger $| \mu_y^S |$, the smaller
$| P_x |$ and $| P_z |$ (and vice versa). The reason is that by increasing $| \mu_y^S |$, we decrease the
the antiferromagnetically coupled $x$ and $z$
components of the magnetization, which are responsible for the inversion symmetry breaking.\cite{BiMnO3}
Thus, there is an unique possibility for controlling the magnetic properties of BiMnO$_3$ by the electric field,
which is directly coupled to the FE polarization: the increase of the polarization should suppress the
FM magnetization. Alternatively, one can control the polarization by the magnetic fields, which is coupled to the
FM magnetization: the increase of the FM magnetization should suppress the polarization. This behavior of
BiMnO$_3$ was predicted theoretically in Ref.~\onlinecite{BiMnO3}. The SCLR theory allows us
to further clarify this behavior on microscopic level.

\section{\label{sec:Summary} Summary and conclusions}

  We have proposed the SCLR method for treating relativistic SO interaction
in the electronic structure calculations. This is the first-order perturbation theory, which also takes into account
the polarization of the electron system by the SO coupling. The method is an efficient alternative to the straightforward
self-consistent solution of Kohn-Sham-like equations with the SO interactions and can be used for the wide class of
magnetic compounds, where the SO interaction is small compared to other parameters of electronic structure.

  The abilities of this method were demonstrated for the solution of effective Hubbard-type model in the
unrestricted HF approximation.
The model itself was derived from the first-principles electronic structure calculations in the Wannier basis and is regarded as
a good starting point for the analysis of magnetic properties of realistic transition-metal oxides
and other strongly correlated systems.

  The SCLR theory brings a substantial improvement over FPA. The latter approach is widely used in the
electronic structure calculations. It is based on the regular perturbation theory with respect to the SO coupling and
totally neglects the effect of electron interactions, which can be also affected by the SO coupling.
The SCLR method becomes especially important when the effective exchange-correlation potential
depends explicitly on the orbital variable,
which is believed to be crucial for treating the orbital magnetization in electronic structure calculations.\cite{OPB,LDAU}

  Moreover,
the main merits of FPA can be easily transferred to SCLR, by replacing the
`bare' SO interaction $\delta \hat{v}^{ext}$ by the `screened' interaction $\delta \hat{v}^{p}$, which takes into account the
polarization of the electron system and, thus, incorporates \textit{all} the contributions in the first order of the SO coupling.
One trivial example is the orbital magnetization, which emerges in the first order of the SO coupling and, therefore,
is well reproduced by the SCLR theory.
Another example is the calculation of antisymmetric DM interactions using the
magnetic force theorem. The DM interactions also emerge in the first-order of the SO coupling and, in principles,
should be accessible by
the perturbation theory for the single-particle energies, as prescribed by the magnetic force theorem.\cite{PRL96}
However, in this perturbation theory, it is also important to include all the contributions in the first order of the SO coupling.
Therefore,
the use of SCLR substantially
improves the description of the DM interactions. The so obtained parameters of spin Hamiltonian appear to be very helpful
in the analysis of complex magnetic structures, which can be realized in realistic materials.

  Another good aspect of SCLR is that it can be combined with variational properties of the total energy. The powerful
$2n$$+$$1$ theorem states in this respect that if one knows the self-consistent potential $\hat{v}$ in the first order of the
SO coupling (or any other perturbation), one
should be able to calculate the corresponding total energy change up to the third order.\cite{2n+1}
This property is very important for compounds with low crystal symmetry. For instance,
one can try to rotate the spin magnetization as the whole and calculate the total energy change caused by the SO coupling.
Typically such calculations give us the MA energy.
Then, what is so special about low-symmetry structures and why is it so important to consider the third-order effects
in this case? Indeed, in uniaxial compounds,
the MA energy is the second-order effect of the SO coupling,\cite{LL,White,Yosida}
and the second order perturbation theory is typically sufficient for reproducing
the corresponding total energy change.\cite{PRB95}
However, when the symmetry is low enough, there will be also the contributions of the DM interactions.
The DM interaction, $\boldsymbol{d}_{ij}$, itself is of the first order of the SO coupling. It produces the canting of spins,
$\delta \boldsymbol{e}_i$ and $\delta \boldsymbol{e}_j$, also in the first order of the SO coupling. Therefore, one can expect
the some additional contributions to the MA energy,
$\boldsymbol{d}_{ij} \cdot [\delta \boldsymbol{e}_i \times \delta \boldsymbol{e}_j]$ in the third order
of the SO interactions, which are captured by the total energy calculations based on the $2n$$+$$1$ theorem.
Other third-order contributions to the total energy are expected from the SI anisotropy terms. These contributions have the
following form:
$\boldsymbol{e}_i^0 \cdot \tensor{\tau}_{ii} \delta \boldsymbol{e}_i +
\delta \boldsymbol{e}_i \cdot \tensor{\tau}_{ii} \boldsymbol{e}_i^0$, which are finite if the tensor
$\tensor{\tau}_{ii}$ has sufficiently low symmetry: since $\boldsymbol{e}_i^0$ and
$\delta \boldsymbol{e}_i$ are orthogonal, the tensor should have non-diagonal
matrix elements.

  Finally, SCLR is a convenient tool for the analysis and interpretation of experimental data and results of
electronic structure calculations with the SO coupling for magnetic materials.
As was discussed above, many applications for such analysis,
which have been earlier developed in the framework of FPA, can be easily adopted for SCLR. In this work, we have demonstrated how
these applications can be used for the analysis of canted magnetic structures in YTiO$_3$ and LaMnO$_3$,
spiral magnetic ordering in BiFeO$_3$, and details of the magnetic inversion breaking in BiMnO$_3$.
The latter application allows us to rationalize several important results,
which were earlier predicted in Ref.~\onlinecite{BiMnO3}. Particularly, the inversion symmetry breaking
by some complex AFM order is typically regarded as the source of the FE activity in improper multiferroics.
In this work, we have argued that the AFM inversion symmetry breaking can not only induce the
FE polarization, but also produce some finite DM interactions, operating across the inversion centers,
which may further lead to the FM canting of spins. Thus, one can expect that in some systems, the AFM inversion symmetry breaking
can be responsible both for the ferroelectricity and the \textit{ferromagnetism}. This is a very unique situation,
which is extremely important
from the viewpoint of
practical realization of the mutual control of electricity and magnetism.
BiMnO$_3$ is the possible candidate, where such a situation could take place.

\appendix*
\section{\label{sec:Appendix}
Energy change due to spin-spiral alignment in the $R3c$ phase of BiFeO$_3$}

  In this appendix, we consider main contributions to the total energy in the case of the spin-spiral alignment
in the noncentrosymmetric $R3c$ phase of BiFeO$_3$.
It is assumed that the spin-spiral alignment is driven by DM interactions.

  Let $\boldsymbol{n}_x$, $\boldsymbol{n}_y$, and $\boldsymbol{n}_z$ be the basis vectors of a
Cartesian coordinate frame, which specify the orientation of the spin spiral.
Namely, it is assumed that the spin spiral lies in the plane spanned by $\boldsymbol{n}_x$ and $\boldsymbol{n}_y$,
while $\boldsymbol{n}_z$ is perpendicular to this plane. Very generally, these three vectors
can be chosen as: $\boldsymbol{n}_x = (-$$\sin \phi, \cos \phi, 0)$,
$\boldsymbol{n}_y = (-$$\cos \phi \cos \theta, -$$\sin \phi \cos \theta, \sin \theta)$, and
$\boldsymbol{n}_z = (\cos \phi \sin \theta, \sin \phi \sin \theta, \cos \theta)$.

  Choosing the phase of the spin spiral such that in the origin
$\boldsymbol{e}_0 = \boldsymbol{n}_x$, the directions of spins
at other sites will be given by
\begin{equation}
\boldsymbol{e}_i =
\boldsymbol{n}_x \cos (\boldsymbol{R}_{0i} \cdot \boldsymbol{q})  + \boldsymbol{n}_y \sin (\boldsymbol{R}_{0i} \cdot \boldsymbol{q}).
\label{eqn:Aedir}
\end{equation}
Then, the energy of DM interactions between NN sites $0$ and $i$, $\boldsymbol{d}_{0i} \cdot [\boldsymbol{e}_0 \times \boldsymbol{e}_i]$,
will be given by
$$(\boldsymbol{d}_{0i} \cdot \boldsymbol{n}_z) \sin (\boldsymbol{R}_{0i} \cdot \boldsymbol{q}) \approx
- (\boldsymbol{d}_{0i} \cdot \boldsymbol{n}_z) (\boldsymbol{R}_{0i} \cdot \delta \boldsymbol{q}),
$$
where $\boldsymbol{q} = \boldsymbol{q}_0 + \delta \boldsymbol{q}$ and
$\boldsymbol{q}_0 = (0,0,2\pi/c)$ corresponds to the collinear G-type AFM alignment, where
$(\boldsymbol{R}_{0i} \cdot \boldsymbol{q}_0) = \pi$ for all NN bonds.
In order to obtain the total energy,
one should sum up the above expression over all six NN bonds
around $0$. In such a construction, the total energy is given per
two Fe sites.
All bonds are connected by the symmetry operations of the space group $R3c$, as explained in
Sec.~\ref{sec:BiFeO3} and Fig.~\ref{fig.BiFeO3}. Therefore, one can write
$$
\delta E_{\rm DM} = - \sum_{g=1}^6 (\hat{S}_g\boldsymbol{d}_{01} \cdot \boldsymbol{n}_z) (\hat{S}_g\boldsymbol{R}_{01} \cdot \delta \boldsymbol{q}),
$$
where $\hat{S}_1 = \hat{E}$ (the unity), $\hat{S}_2 = \hat{C}^{3+}_z$, $\hat{S}_3 = \hat{C}^{3-}_z$, $\hat{S}_4 = -$$\hat{m}_y$,
$\hat{S}_5 = -$$\hat{C}^{3+}_z \hat{m}_y$, and $\hat{S}_6 = -$$\hat{C}^{3-}_z \hat{m}_y$. Then, using the explicit matrix form for
$$
\hat{C}^{3\pm}_z =
\left(
\begin{array}{ccc}
-1/2           & \pm \sqrt{3}/2 & 0 \\
\mp \sqrt{3}/2 & -1/2           & 0 \\
0              & 0              & 1 \\
\end{array}
\right),
\quad
\hat{m}_y =
\left(
\begin{array}{ccc}
1              & \phantom{-}0              & 0 \\
0              & -1             & 0 \\
0              & \phantom{-}0              & 1 \\
\end{array}
\right),
$$
and $\hat{C}^2_y = -$$\hat{m}_y$, which acts on the vectors of DM interactions,
one can obtain the following expression:
\begin{equation}
\delta E_{\rm DM} = 3 [\boldsymbol{n}_z \times \delta \boldsymbol{q}]_z [\boldsymbol{R} \times \boldsymbol{d}]_z,
\label{eqn:ADM}
\end{equation}
where $\boldsymbol{R} = (R_x,R_y,R_z)$ and $\boldsymbol{d} = (d_x,d_y,d_z)$ refer to the bond $0$-$1$ or to any other NN bond.
Thus, $E_{\rm DM}$ in the spin-spiral state does not depend on $d_z$. Moreover, the energy gain due to
DM interactions is maximal when $\delta \boldsymbol{q} \perp \boldsymbol{n}_z$.
By choosing the bond with $\boldsymbol{R} = (a,0,-$$c/2)$ (see Fig.~\ref{fig.BiFeO3}), one obtains
$$
\delta E_{\rm DM} = 3a d_y [\boldsymbol{n}_z \times \delta \boldsymbol{q}]_z .
$$

  By applying the same strategy, the energy loss due to NN isotropic exchange interactions
can be evaluated as (also per two Fe sites)
$$
\delta E_{\rm H} = - J \sum_{g=1}^6 \cos (\hat{S}_g\boldsymbol{R}_{01} \cdot \boldsymbol{q}) -6J \approx
-\frac{J}{2} \sum_{g=1}^6 (\hat{S}_g\boldsymbol{R}_{01} \cdot \delta \boldsymbol{q})^2,
$$
which yields:
$$
\delta E_{\rm H} = - \frac{3J}{2} (R_x^2 + R_y^2) \left( (\delta q_x)^2 + (\delta q_y)^2 \right) + 3J R_z^2 (\delta q_z)^2.
$$
Since, $R_x^2 + R_y^2 = a^2$ and $R_z^2 = c^2/4$, this
expression can be further transformed to
$$
\delta E_{\rm H} = - \frac{3a^2J}{2} \left( (\delta q_x)^2 + (\delta q_y)^2 \right) - \frac{3c^2J}{4}  (\delta q_z)^2.
$$
Similar expression for the next-NN interactions in the $xy$-plane is obtained by noting that, in this case,
$R_x^2 + R_y^2 = 3a^2$, $R_z = 0$, and $\boldsymbol{q}_0$ corresponds to the FM coupling between these
next-NN spins. This yields
$$
\delta E_{\rm H}' = \frac{9a^2J'}{2} \left( (\delta q_x)^2 + (\delta q_y)^2 \right).
$$

  For the direction of spin $\boldsymbol{e}_i$,
given by Eq.~(\ref{eqn:Aedir}), the SI anisotropy energy has the following form:
$$
(\boldsymbol{e}_i \cdot \tensor{\tau} \boldsymbol{e}_i) = -\frac{1}{2} \tau_{zz} +
\frac{3}{2} \tau_{zz} \sin^2 \theta \sin^2 (\boldsymbol{R}_{0i} \cdot \boldsymbol{q}),
$$
where we have used the fact that $\tensor{\tau}$ is the diagonal tensor with the matrix elements
$\tau_{xx} = \tau_{yy} = -$$\frac{1}{2}\tau_{zz}$. Noting that $\boldsymbol{R}_{0i} \cdot \boldsymbol{q}_0 = n \pi$
($n$ being an integer number), the above expression can be further transformed to
$$
(\boldsymbol{e}_i \cdot \tensor{\tau} \boldsymbol{e}_i) = -\frac{1}{2} \tau_{zz} +
\frac{3}{2} \tau_{zz} \sin^2 \theta \sin^2 (\boldsymbol{R}_{0i} \cdot \delta \boldsymbol{q}).
$$
Then, the change of the SI anisotropy energy
is obtained by averaging the second term over all possible angles between
$\boldsymbol{R}_{0i}$ and $\delta \boldsymbol{q}$. For the homogeneous spin spiral, the phase
$\boldsymbol{R}_{0i} \cdot \delta \boldsymbol{q}$ changes by an equal amount between neighboring lattice points. Moreover, for
small $\delta \boldsymbol{q}$, the summation over discrete angles can be replaced by integration,
which yields for the change of the SI anisotropy energy (per two Fe sites):
$$
\delta E_{\rm SI} = \frac{3}{2} \tau_{zz} \sin^2 \theta.
$$
Thus, $\delta E_{\rm SI}$ depends only on the orientation $\theta$ of the spin spiral relative to the
anisotropy axis. However, it
does not depend on $\delta \boldsymbol{q}$, in agreement with the previous finding.\cite{Sosnowska}

  Nevertheless, we would like to emphasize that this expression is
valid only for the \textit{homogeneous} spin-spiral state, which \textit{was enforced} in the preset analysis.
In a more general case, the SI anisotropy is responsible for the ``bunching'' of magnetic moments,\cite{REbunching}
which leads to the deformation of the spin-spiral state. In the \textit{deformed} spin-spiral state,
$\delta E_{\rm H}$, $\delta E_{\rm DM}$, and $\delta E_{\rm SI}$ can reveal a
different $\boldsymbol{q}$-dependence, because all these quantities will depend
on the additional phases of magnetic moments, which are acquired due to the bunching. Thus,
in more general magnetic structures, the value of $\delta \boldsymbol{q}$ can be also
controlled by the SI anisotropy term.


\begin{thebibliography}{99}
\bibitem{remark1}
Note, that in order to be consistent with our definition, Eq.~(\ref{eqn:spinE}),
all experimental parameters of spin Hamiltonian for BiFeO$_3$ should be multiplied by $S^2 = (5/2)^2$.
Then, the values isotropic interactions, reported in Ref.~\onlinecite{BiFeO3Jeong}, correspond to
$|J| = 27.38$ meV and $|J'| = 0.94$ meV,
while the values reported in Ref.~\onlinecite{BiFeO3Matsuda}
correspond to $|J| = 40.50$ meV and $|J'| = 1.81$ meV.
Thus, our parameters, $|J| = 37.27$ meV and $|J'| = 1.60$ meV, are roughly in-between
these two sets of experimental data. Tne effective DM interaction,
reported in Ref.~\onlinecite{BiFeO3Matsuda},
corresponds to $|d_y| = 1.014$ meV in our definition.
The experimental value of $|d_y|$ reported in Ref.~\onlinecite{BiFeO3Jeong},
in addition to $S^2 = (5/2)^2$, whould be multiplied by two, which yields
$|d_y| = 1.338$ meV. The single-ion anisotropy energy was estimated in Ref.~\onlinecite{BiFeO3Matsuda}.
It corresponds to $E_{\rm SI} = 0.043 \pm 0.004$ meV in our definition.
\end{thebibliography}
\end{document}